\documentclass[traditabstract]{aa}
\usepackage{txfonts,natbib,graphicx}
\usepackage{lscape}
\bibpunct{(}{)}{;}{a}{}{,}

\newcommand{\mxb}{4U~1705--44}

\newcommand{\xmm}{{\it XMM-Newton}}
\newcommand{\beppo}{{\it BeppoSAX}}
\newcommand{\suzaku}{{\it Suzaku}}
\newcommand{\chandra}{{\it Chandra}}
\newcommand{\rxte}{{\it RXTE}}
\newcommand{\integral}{{\it INTEGRAL}}

\def\ltsima{$\; \buildrel < \over \sim \;$}
\def\simlt{\lower.5ex\hbox{\ltsima}}
\def\gtsima{$\; \buildrel > \over \sim \;$}
\def\simgt{\lower.5ex\hbox{\gtsima}}

\begin{document}

\title{Testing reflection features in \mxb\ with \xmm, \beppo\ and \rxte\ in the hard and soft state
}
\author{E.~Egron$^{1}$\thanks{mail to elise.egron@dsf.unica.it} \and 
  T.~Di Salvo$^{2}$ \and S. Motta$^{3,4}$ \and L.~Burderi$^{1}$ \and A.~Papitto$^{5}$ \and 
R. Duro$^{6}$
\and A.~D'A\`i$^{2}$ 
\and A.~Riggio$^{1,7}$ 
\and T.~Belloni$^{3}$ 
\and R.~Iaria$^{2}$
\and N.~R.~Robba$^{2}$
\and S.~Piraino$^{8,9}$ 
\and A.~Santangelo$^{9}$ 
}
\institute{Dipartimento di Fisica, Universit\`a degli Studi di Cagliari, 
SP Monserrato-Sestu, KM 0.7,  09042 Monserrato, Italy
\and Dipartimento di Fisica, Universit\`a di Palermo, 
via Archirafi 36, 90123 Palermo, Italy 
\and INAF-Osservatorio Astronomico di Brera, Via E. Bianchi 46, I-23807 Merate (LC), Italy 
\and Universit\`a dell' Insubria, Via Valleggio 11, I-22100 Como, Italy
\and Institut de Cie\`ncies de l'Espai (IEEC-CSIC), Campus UAB, Fac. de Cie\`ncies, Torre C5, parell, 2a planta, 08193 Barcelona, Spain
\and Dr. Karl Remeis-Sternwarte and Erlangen Centre for Astroparticle Physics,
Friedrich-Alexander-Universit\"at Erlangen-N\"urnberg, Sternwartstra$\ss$e 7, 
96049 Bamberg, Germany
\and INAF - Osservatorio Astronomico di Cagliari, Poggio dei Pini, Strada 54, 
09012 Capoterra (CA), Italy
\and INAF-IASF di Palermo, via Ugo La Malfa 153, 90146 Palermo, Italy
\and Institut f\"ur Astronomie und Astrophysik, Kepler Center for Astro and Particle Physics, Sand 1, 72076 T\"ubingen Germany}

\abstract
{We use data of the bright atoll source \mxb\ taken with \xmm, \beppo\ and \rxte\ both in 
the hard and in the soft state to perform a self-consistent study of the 
reflection component in this source.
Although the data from these X-ray observatories are not simultaneous, the
spectral decomposition is shown to be consistent among the different observations, when the source flux is similar.
We therefore select observations performed at similar flux levels in the 
hard and soft state in order to study the spectral shape in these two states
in a broad band (0.1--200 keV) energy range, with good energy resolution,
and using self-consistent reflection models. These reflection 
models provide a good fit for the X-ray spectrum both in the hard and 
in the soft state in the whole spectral range. We discuss the differences in the main spectral parameters we find in the hard and the soft state, respectively, providing evidence that
the inner radius of the optically thick disk slightly recedes in the hard state.}

\titlerunning{Testing reflection features in \mxb\ with \xmm, \beppo\ and \rxte\ in the hard and soft state}
\authorrunning{Egron et al.}

\maketitle

\section{Introduction}

The X-ray emission in low mass X-ray binaries (LMXBs) 
comes from the gravitational potential energy released from accretion processes onto black holes or neutron stars. The X-ray spectrum is generally well-described by a soft-thermal component such as a blackbody or a multicolor-disk blackbody, originating from the accretion disk, and a hard X-ray component which usually dominates the spectrum. This hard component can be fitted either by a power-law with a high energy cutoff or a unsaturated Comptonization spectrum, when the source is in the so-called hard state, 
or by a blackbody or a saturated Comptonization spectrum, when the source is in the soft state, where the temperature of the photons is very similar to the electron temperature.
The hard component is generally explained in terms of inverse Compton scattering, where soft thermal photons get Compton up-scattered by hot electrons forming a corona or a boundary layer between the neutron star surface and the accretion disk,
or forming the base of a jet \citep[at least the hard state,][]{Markoff_04, Markoff_05}. 

In addition to this, a broad iron line (associated to the Fe K$\alpha$ emission) is often detected in the spectra of X-ray binaries. However, the nature of its large width and the (a)symmetric profile of the line are still debated.
In analogy with systems containing stellar mass or supermassive black holes, which show a remarkably similar phenomenology \citep[e.g.][]{Martocchia_96, Walton_12}, it can be produced by reflection in the accretion disk \citep{Reynolds_03,Fabian_05,Matt_06}, or it 
can arise from an accretion disk corona \citep{Kallman_89, Vrtilek_93}.

The first scenario implies that the iron line is produced in the inner part of the accretion disk. Hard X-rays coming from the corona or from the base of the jet irradiate the relatively cold accretion disk. 
As a consequence, a broad and asymmetric line is expected due to Doppler and relativistic effects produced close to the compact object. 
Depending on the ionization state of the disk, this leads to the emission of several emission lines and absorption edges that are more or less strong in the spectrum, depending on the relative abundance of the corresponding ion and/or its fluorescence yield \citep{Kaastra_93}. The Fe K$\alpha$ line at 6.4$-$7 keV is the most prominent feature \citep{Fabian_00}. It results in
a fluorescent line at 6.40 keV from Fe I-XVII, or recombination lines at $6.67-6.70$ keV and $6.95-6.97$ keV associated with highly ionized species of Fe XXV (He-like) and Fe XXVI (H-like), respectively.
In this scenario other reflection signatures are also expected, such as the emission and/or absorption from several elements at lower energy, and a Compton reflection hump at higher energy \citep{george_91, Ballantyne_01,Ross_07} if the X-ray continuum spectrum is sufficiently hard. \citet{dai_09}, \citet{Iaria_09}, \citet{disalvo_09} and \citet{piraino_12} detected emission lines in GX 340+0, GX 349+2, \mxb\ and GX 3+1 at 2.6, 3.31 and 3.90 keV associated with S XVI, Ar XVIII and Ca XIX, respectively, and an absorption edge from ionized iron. Moreover a reflection hump peaking at nearly 30 keV has been found in several sources, such as \mxb\ \citep{Fiocchi_07}.
To confirm the common origin of these features, it is important to use self-consistent models which include all the reflection components in order to discriminate among the proposed production mechanisms.

Alternative scenarios explain the line broadening by Compton scattering if the line is emitted in the inner parts of a moderately optical thick accretion disk corona (ADC), formed by evaporation of the outer layers of the disk \citep{White_82}. Another possibility is that the line profile is red-skewed due to Compton up-scattering by a narrow wind shell launched at mildly relativistic velocities at some disk radii where the local radiation force exceeds the local disk gravity \citep{Titarchuk_09}.


Due to the good energy resolution capabilities of \chandra,\ and to the large effective
area of \xmm\ and \suzaku,\ the number of significant detections of these lines is increasing, giving the possibility to study their profile in more detail.
Recently, two studies were performed by \citet{Ng_10} and \citet{Cackett_10} in order to investigate the nature of the iron line using a large sample of neutron star LMXBs spectra.
Using \xmm\ observations of 16 neutron star LMXBs, \citet{Ng_10} conclude that there is no statistical evidence that the iron line profile is asymmetric and they propose that 
the observed large width of the line is caused by
Compton scattering in the corona. It should be noted that, in order to eliminate the effects of photon pile-up in the \xmm\ data, \citet{Ng_10} decided to reject up to 90\% of the source photons for the brightest sources. Conversely, by studying 10 neutron star LMXBs with \suzaku\ (which is less affected by photon pile-up because of a broad point spread function) and \xmm,\ \citet{Cackett_10} confirm that the Fe line is asymmetric, relativistic, and produced by reflection in the inner part of the accretion disk. This result was achieved also comparing CCD-based spectra from \suzaku\ with Fe K line profiles from archival data taken with gas-based spectrometers \citet{Cackett_12}. In general, they found a good consistency between the gas-based line profiles from EXOSAT, BeppoSAX, RXTE, and the CCD data from \suzaku,\ demonstrating that the broad profiles seen are intrinsic to the line and not broad due to instrumental issues.

\subsection{The case of 4U 1705-44}

\mxb\ is a low mass X-ray binary system containing a weakly magnetized neutron star. In such systems the accretion disk can extend down to the neutron star surface. This implies a similar configuration with respect to that envisaged for accreting black holes, since the radius of the neutron star is close to the size of the innermost stable circular orbit (ISCO) of matter around a black hole.
Similarities have been observed between these systems \citep[e.g.,][and references therein]{Maccarone_12}, suggesting common physical processes producing X-ray emission and similar properties of the accretion flow around the compact object. Clear observational evidences exist that a neutron star has a solid surface, such as the observation of type I X-ray bursts, which are thermonuclear explosions in the surface layers of the neutron star, or of coherent pulsations, resulting from the magnetic field anchored on the neutron star surface. However not all systems containing a neutron star show one or more of these characteristics \citep{Done_07}.

\mxb\ is a persistent bright source that shows type I X-ray burst \citep{Langmeier_87} and kHz quasi-periodic oscillations \citep{Ford_98, Olive_03}. From some bursts that show photospheric radius expansion, its distance is estimated at 7.4 kpc, toward the Galactic ridge \citep{Forman_78, Galloway_08}.		

LMXBs containing a neutron star are divided into two categories \citep{Hasinger_89} according to the path the source describes in the X-ray color-color diagram (CD) or hardness-intensity diagram (HID): atoll sources (C-like, low luminosity, about 0.001$-$0.5 $ L_\mathrm{Edd}$)  or Z-sources (Z-like, high luminosity, close to the Eddigton limit). \mxb\ is classified as an atoll source. 
In the classical CD pattern of these sources, two branches are usually distinguished: the island and the banana.
In the island branch, the source presents a low count rate, probably associated to a low mass accretion rate, and the source is in the low/hard state. In this state, it is thought that the accretion disk is truncated relatively far from the compact object. This results in a very hot corona and in a hard spectrum.
On the other hand, when the source is in the banana branch, the accretion disk approaches the compact object, the temperature of the corona decreases due to Compton cooling, and the source is in the high/soft state (high is relative to the X-ray flux, and soft to the spectrum).
The mass accretion rate therefore is expected to increase from the island to the banana branch.
The Z-sources usually emit persistently at high luminosity and are thought to have a larger magnetic field and/or higher mass accretion rate than atoll sources.

\mxb\ has recently been studied thoroughly with the particular objective to understand the origin of its iron line. A broad Fe line was clearly visible with \chandra\ \citep{disalvo_05}, confirming previous studies performed at low energy resolution by \citet{White_86} and \citet{Barret_02}. The HETG on board \chandra\ excluded that the large width of the line could be caused by the blending of lines from iron at different ionization states, although it was not possible to discriminate between the two possible origins (relativistic smearing or Compton broadening) of the line width. The broad band \beppo\ spectrum of \mxb\ taken during a soft state again showed a broad iron line that could be well fitted by a \textsc{diskline} component with very reasonable smearing parameters \citep{Piraino_07}, and the \integral\ spectrum showed clear evidence of a Compton reflection hump \citep{Fiocchi_07}, as well as the \beppo\ spectrum taken during the hard state (Piraino et al. in preparation). 

The observations with \xmm\ and \suzaku\ were essential to study the iron line profile in detail because of their large effective area and good energy resolution. Indeed, \citet{Reis_09} found a
broad (skewed) and asymmetric Fe K$\alpha$ emission line during a \suzaku\ observation of \mxb.\ 
Both Compton broadening and relativistic smearing were necessary to account for the large width
of the iron line. The best fit value of the inner radius was R$_{in} \simeq 10.5$~R$_g$, with
R$_g$ being gravitational radius (R$_g$ $= GM/c^{2}$), and the corresponding inclination angle
was $i \sim 30^\circ$. 
Using the observations from \beppo\ and \suzaku,\ \citet{lin_10} obtained constraints on the
broad-band spectrum of \mxb\ and found that the strength of the Fe line correlates well with the
boundary layer emission in the soft state, while the Fe line is probably due to illumination of the
accretion disk by the strong Comptonization emission in the hard state.
An \xmm\ observation performed in 2008 when the source was in a soft state confirmed the results obtained with \suzaku,\ showing a high-statistics Fe line profile and spectrum consistent with a disk-reflection scenario \citep{disalvo_09,dai_10}. The best fit value of the inner disk radius was R$_{in} \simeq 14$~R$_g$ and the corresponding inclination angle and index of the emissivity law profile were $i \sim 39^\circ$ and $-2.3$ respectively, in agreement with the \suzaku\ results. In addition, the \xmm\ spectrum showed the presence of several emission lines from elements lighter than iron, such as S, Ar, and Ca at 2.62 keV, 3.31 keV and 3.90 keV, respectively, and an absorption edge at 8.5 keV probably from ionized iron. All these features were broad and consistent with being produced in the same region where the iron line was produced. 

Here we present the spectral analysis of \mxb\ using data from \xmm,\ \beppo,\ and \rxte\ when the source was in the hard state and in the soft state. Using the same spectral models to describe the source spectrum in both states, we aim at highlighting the differences in the spectral parameters finding that the accretion disk is truncated further (but not far) from the compact object in the hard state. Since in \mxb\ several reflection features have been detected, we use self-consistent reflection models to fit the broad-band, multi-mission, X-ray spectrum of this source. We also discuss possible effects due to pile-up distortion in \xmm\ spectrum during the high-luminosity, soft state.

\section{Observation and data reduction}

\begin{figure}[t]
\includegraphics[width=9.3cm]{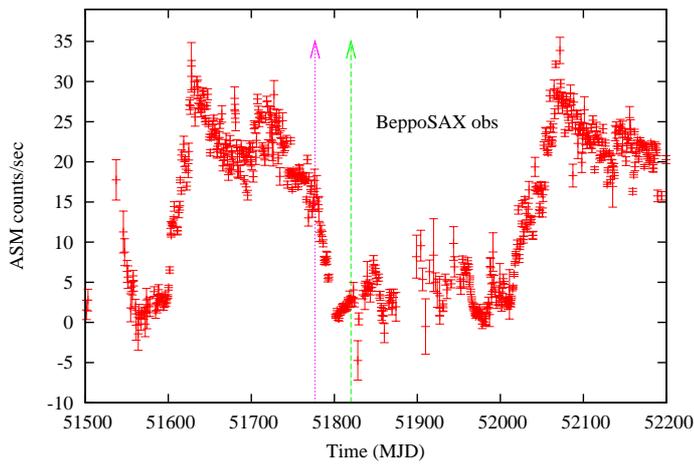}
\caption{Light curve of 4U 1705-44 obtained with RXTE/ASM from November 1999 to October 2001. The two arrows show the observations performed by \beppo\ in August and October 2000.}
\label{fig:lc_beppo}
\end{figure}

\begin{figure}[t]
\includegraphics[width=9.3cm]{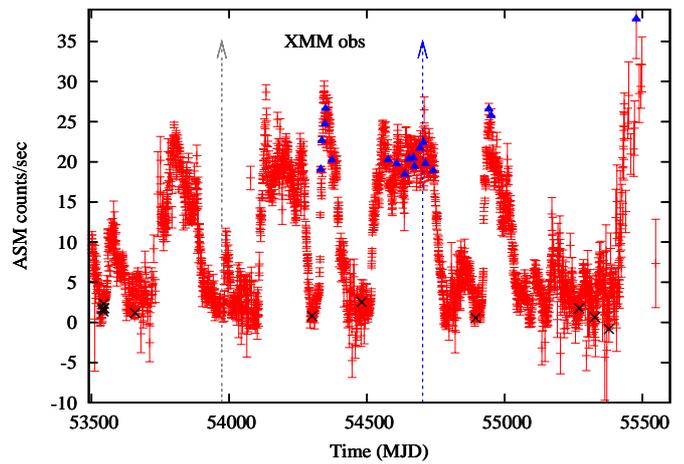}
\caption{RXTE/ASM light curve of 4U 1705-44 from May 2005 to November 2010. The arrows mark the two \xmm\ observations performed in August 2006 and 2008. The black crosses and the blue triangles represent the \rxte\ observations selected in the hard and soft state, respectively.}
\label{fig:lc_xmm}
\end{figure}

The light curve produced from the All-Sky Monitor on-board \rxte\ allows to follow the evolution of the source flux for a period of $\sim 16$ years. The source shows clear spectral transitions, from the hard (3 counts/s) to the soft state (25 counts/s).  
A total of four observations were performed with \beppo\  (see Fig.~\ref{fig:lc_beppo}) and \xmm\  (see Fig.~\ref{fig:lc_xmm}) when the source was in the soft and hard state.
The spectral transitions are associated with variations in the X-ray flux, likely proportional to the accretion rate.
We select \rxte\ observations, both in the hard and soft state, as shown in 
Fig.~\ref{fig:lc_xmm} .
In this way, the joint spectra from the three satellites cover the full $0.1-200$ keV energy range.
We explain in detail our methods in the following. 

\subsection{XMM-Newton}

\mxb\ was observed twice with \xmm. The first observation was performed on 2006 August 26 for an effective exposure of 34.72 ks, when the source was in the hard state. The corresponding RXTE/ASM count rate  was 1 c/s. The second time, the source was observed during the soft state (Target of Opportunity), on 2008 August 24 for a total on-source observing time of 45.17 ks \citep[see][for more details on this observation]{disalvo_09}. The RXTE/ASM was 19 c/s during this observation.

During both observations, the European Photon Imaging Camera pn (EPIC-pn; Struder et al. 2001) and the Reflection Grating Spectrometers (RGS1 and RGS2; den Herder et al. 2001) were used to observe the source. The EPIC-pn camera operated in timing mode to minimize photon pile-up and telemetry overload that may occur at high count rates, with a thick filter in place which further reduced the number of low-energy photons. In timing mode only the central CCD is read out with a time resolution of 30\,$\mu$s. This provides a one-dimensional image of the source with the second spatial dimension being replaced by timing information.
The EPIC-MOS and the optical Monitor were off during the second observation in order to avoid telemetry drop-outs in the EPIC-pn.

We produced a calibrated photon event file using the SAS\footnote{The \xmm\ data were processed using the Science Analysis Software v.10, following the \xmm\ ABC guide.} processing tool \textsc{epproc}. Before extracting the  spectra we checked for contamination from background solar flares by producing a light curve in the energy range 10$-$12 keV. No solar flare was registered. However, during the first observation (corresponding to the hard state), the EPIC-pn camera registered a type-I X-ray burst at about 11 ks after the start of the observation. We applied temporal filters by creating a good time interval (GTI) file with the task \textsc{tabgtigen} in order to remove the burst.

We used the task \textsc{epfast} to correct rate-dependent CTI effects in the event list. The source spectra were extracted from a rectangular area covering all the pixels in the Y direction and centered on the brightest RAWX column (36 in the hard state, and 38 in the soft state) with a width of 16 pixels, which corresponds to 65.6 arcsec. We selected only events with $\mathrm{PATTERN} \leq 4$ (single and double pixel events) and FLAG=0 as a standard procedure to eliminate spurious events.  
We extracted the background spectra from a box similar to the one used to extract the source photons but in a region away from the source ($\mathrm{RAWX}=47-63$ and $\mathrm{RAWX}=4-12$ in the hard and soft state, respectively).
During the first observation, the average count rate registered by EPIC-pn CCDs is around 48 c/s, and 19 c/s in the 2.4$-$11 keV range, excluding the burst interval. In the soft state, the  count rate is much higher; the mean count rate is around 770 count/s, and 425 count/s in the 2.4$-$11 keV range, slightly increasing in time (by 5\%).
We also checked for the presence of pile-up using the 
task \textsc{epatplot}. While the first \xmm\ observation of the source taken in the hard state does not show any significant pile-up, a few percent pile-up affects the second observation taken during a soft state.
We discuss the pile-up issue concerning the soft state in the next section.
We grouped the EPIC-pn energy channels by a factor of 4 in order to avoid
oversampling of the energy resolution bin of the instrument.

The two RGS units were set in the standard spectroscopy mode.The RGS data were processed 
using the \textsc{rgsproc} pipeline to produce
calibrated event list files, spectra and response matrices. 
The RGS data were rebinned to have at least 25 counts per energy channel. 

\subsection{BeppoSAX}

\beppo\ performed two observations of \mxb\ in August and October 2000, for a total on-source observing time of 43.5 ks and 47 ks respectively. The count rate registered by  the RXTE/ASM associated to these observations was 18 c/s and 3 c/s respectively.

The four \beppo\ narrow field instruments were on during both the observations. The Low Energy Concentrator Spectrometer (LECS, 0.1$-$4 keV; Parmar et al. 1997) and the Medium Energy Concentrator Spectrometer (MECS, 1$-$10 keV; Boella et al. 1997) data were extracted in circular regions centered on the source position using radii of 8' and 4', respectively, corresponding to 95\% of the source flux. Identical circular regions were used in blank field observations to produce the background spectra. The background spectra of the High Pressure Proportional Gas Scintillation Counter (HPGSPC, 8$-$50 keV; Manzo et al. 1997) and of the Phoswich Detection System (PDS, 15$-$200 keV; Frontera et al. 1997) were produced from Dark Earth data and during off-source intervals, respectively. The HPGSPC and PDS spectra were grouped using a logarithmic grid.

We did not use data from PDS during the first observation, which corresponds to the soft state, in order to avoid extra complication of the spectral fit caused by the presence of a hard (power-law) spectral component \citep[see][]{Piraino_07}.
During the low/hard observation, 6 X-ray bursts were removed from the data.

\subsection{RXTE}

\begin{figure}[t]
\includegraphics[height=6.5cm]{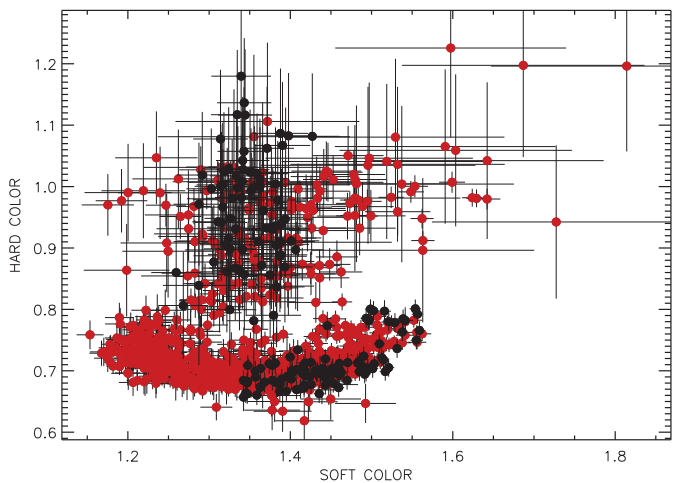}
\caption{Color-color diagrams (CDs) of \mxb\ produced by \rxte/PCA and \xmm/EPIC-pn, in red and black respectively. The CDs were Crab normalized, and an additional correction was needed to precisely match the two diagrams. The \rxte\ observations have been selected in such way that they match with those of \xmm,\ in the hard and soft state.}
\label{fig:ccd}
\end{figure}

There was no simultaneous observation performed by \rxte\ during the \xmm\ and \beppo\ observations. However, the archive provides us with hundreds of \rxte\ observations of \mxb\ since \rxte\ was launched. Therefore, we considered all the observations of \mxb\ collected since 2000 May (corresponding to the 5th epoch of RXTE\footnote{We only considered data from the 5th epoch to avoid fluctuations due to the differences in the instrument gain that can be observed in data coming from different epochs.}) and selected \rxte\ observations during which the source showed the same spectral state of the two \xmm\ observations mentioned above. To do so we produced a color-color diagram (CD) from all the \rxte\ observations and a time-resolved CD from the two \xmm\ observations (we separated the \xmm\ observations in intervals 512 s long, and for each of them, we produced a spectrum and measured the colors) using the same energy bands for the two instruments (i.e. $2.47-3.68$ keV and $3.68-5.31$ keV for the soft color and, $5.31-7.76$ keV, $7.76-10.22 $ keV for the hard color). Since from the \rxte\ data it is clear that the shape of the CD of \mxb\ remains constant during the period considered, we could directly compare the CD coming from the two instruments. 
The CDs were normalized to the Crab colors, but an additional correction was necessary to precisely match the two diagrams, to take into account the differences in the gain of the two instruments. Fig.~\ref{fig:ccd} shows the \xmm\ CD corresponding to the two observations performed in the hard and in the soft state superposed to the \rxte\ CD. Starting from the two CDs we selected the \rxte\ observations that matched the \xmm\ observations. The detail of the \rxte\ observations selected is given in Table~\ref{tabps:rxtedet}. 

The \rxte\ data were obtained in several simultaneous modes. 
STANDARD 2 and STANDARD mode for the PCA and HEXTE instruments, respectively, were used to create background
and dead time corrected spectra. We extracted energy spectra from PCA and HEXTE for
each observation using the standard \rxte\ software within HEAsoft v.6.9 following
the standard procedure described in the \rxte\ cookbook
to produce source and background spectra as well as response matrices. 
Only Proportional Counter Unit 2 from the PCA was used since only this unit was on
during all the observations.
As regards HEXTE data, we used only data coming from HEXTE/Cluster B, which were correctly working in our period of interest. 

Then we produced RXTE/PCA and RXTE/HEXTE spectra for each \rxte\ observation and we averaged them to obtain a PCA+HEXTE spectrum in the island state (matching \xmm\ observation made in 2006 August 26) and a PCA spectrum in the banana-state (matching \xmm\ observation made in 2008 August 24). In the soft state we did not use RXTE/HEXTE data due to the lack of counts in the HEXTE working enery range ($20-200$ keV). 
A systematic error of 0.6\% was added to the PCA averaged spectra to account for residual uncertainties in the instrument calibration\footnote{http://www.universe.nasa.gov/xrays/programs/rxte/pca/doc/rmf/pcarmf-11.7 for a detailed discussion on the PCA calibration issues.}. 

\section{Spectral analysis}

Data were fitted by using Xspec \citep{arnaud_96} v.12.6. All uncertainties are given at the 90\% confidence level ($\Delta \chi^2 = 2.706 $).

The data analysis of the XMM/EPIC-pn spectrum was restricted to 2.4$-$11 keV to exclude the region around the detector Si K-edge (1.8 keV) and the mirror Au M-edge (2.3 keV) that could affect our analysis. This problem was already noticed for the EPIC-pn 
observations performed in timing mode \citep[e.g.,][]{dai_10,papitto_10,Egron_11}.
The energy bands used for the other instruments are: 0.3$-$4 keV for the LECS, 1.8$-$10 keV for the MECS, 7$-$34 keV for HPGSPC, 15$-$200 keV for PDS on-board \beppo,\ and 4$-$22 keV for PCA and 15$-$100 keV for HEXTE on-board \rxte.\

The following sections concern the analysis of data associated to the soft and to the hard state of \mxb.\ In the first one, we investigate the pile-up effects on the \xmm\ data of the soft state; we demonstrate that the iron line is always broad and gives consistent spectral parameters independent of the extraction region. Then we apply reflection models to fit the \xmm\ data in this state, and finally we include the \beppo\ and \rxte\ data to extend the analysis on the broad-band, $0.1-200$ keV, energy range. In the second one, we apply the same continuum and reflection models to the \xmm,\ \beppo\ and \rxte\ data in the hard state, in order to evaluate which spectral parameters change from one state to the other.

\section{Soft state}

\subsection{Pile-up in \xmm\ /EPIC-pn data?}

Pile-up is an important issue for CCD data and may affect the spectral results. It occurs when more than one X-ray photon hits the same pixel or an adjacent one in the same read-out frame.
If this happens the CCD will be unable to resolve the individual photon events and instead record a single event with an energy that is roughly the sum of the individual event energies. It results in a shift of the photons to higher energy, which produces an energy-dependent distortion of the spectrum.
It is possible to assess the pile-up effects by checking the fraction of singles, doubles, triples and quadruples events (depending on how many pixels are involved) using the task \textsc{epatplot} (see Appendix). 

\citet{disalvo_09}, \citet{dai_10} and \citet{Ng_10} studied the iron line in \mxb\ using the same  \xmm\ data in the soft state, but considering different extraction regions for the spectra and therefore accepting different  pile-up fraction in their spectral analysis. 
\citet{disalvo_09} considered that pile-up effects on spectral results were negligible and hence decided to keep all the central columns of the CCD, whereas \citet{dai_10} excluded the brightest CCD column, and \citet{Ng_10} excluded 7 central columns (corresponding to $\sim$ 90\% of the source counts). In all these cases, the iron line detected in the pn spectra of \mxb\ 
remained consistently broad, with a Gaussian sigma $\sim 0.3-0.4$ keV. However, the conclusions coming from the spectral analysis differed. While \citet{disalvo_09} and \citet{dai_10} deduced that the iron line is clearly asymmetric and compatible with a relativistic line, using the \textsc{diskline}
\citep{fabian_89} and the reflection model \textsc{refbb} \citep{Ballantyne_04}, respectively, \citet{Ng_10} found that the iron line can be fitted equally well by a gaussian or using the \textsc{laor} model \citep{laor_91}, concluding that there was no statistical evidence for an asymmetry of the line profile. It is thus important to assess the effects of pile-up in the EPIC-pn spectrum of \mxb\ in the soft state, which shows the highest S/N ratio iron line profile ever detected to date in a neutron star LMXB. 

\begin{figure}[t]
\includegraphics[height=9.1cm,angle=-90]{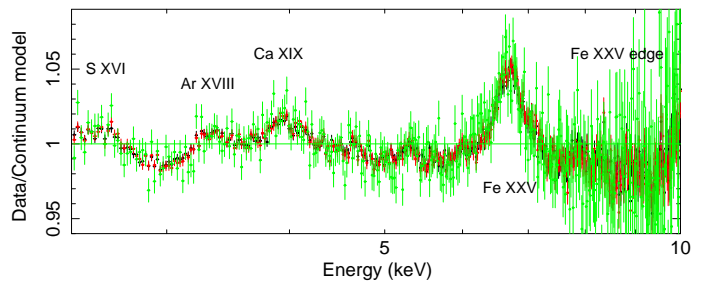}
\caption{Comparison of the 
ratios of the data to the best-fit continuum 
considering all the rows of the CCD (black; Di Salvo et al. 2009), without the brightest CCD column (red; D'Ai et al. 2010), and excluding 7 central columns (green; Ng et al. 2010). The continuum model consists of a blackbody and a Comptonization model (comptt) modified by photoelectric absorption (phabs).}
\label{fig:res_3}
\end{figure}

In order to evaluate the pile-up effects on the \xmm/pn spectrum, we apply an empirical model, similar to the one used in \citet{disalvo_09}, to compare the results obtained when we consider different extraction regions in the CCD. In particular, we consider the extraction region described in Sec. 2, where we exclude from 0 up to 7 central brightest columns before extracting the pn spectrum. Fig.~\ref{fig:res_3} shows the ratios of the data to the continuum obtained for three cases: excluding 0, 1, and 7 central brightest columns. The continuum model consists of a blackbody and a Comptonization component (\textsc{compTT}; Titarchuk 1994), modified at low energy by the photoelectric absorption (\textsc{phabs}; photoelectric cross-sections of \citet{balucinska_98} with a new He cross-section based on \citet{yan_98} and standard abundances of \citet{anders_89}).
This model is often used for atoll sources and gives a good fit for the continuum of this source \citep{Barret_02, disalvo_05, Piraino_07, disalvo_09}. Three emission lines are visible at low energy, at 2.62 keV, 3.31 keV and 3.90 keV, identified by \citet{disalvo_09} as highly ionized elements corresponding to S XVI, Ar XVIII and Ca XIX, respectively. In addition to these lines, an iron emission line and an absorption edge are present. The detection of the iron line is at about 10 $\sigma$ above the continuum when all the columns are considered whereas it becomes only 5 $\sigma$ when 7 columns are removed. In the three cases, the iron line appears broad and the shape is very similar. To fit these residuals we add four gaussians and an edge to the continuum model, all modified by the same relativistic blurring (modelled with \textsc{rdblur}, the \textsc{diskline} kernel, in Xspec) component to take into account the relativistic and/or Doppler effects produced by the motion in the inner disk close to the compact object. 
This model describes the relativistic effects due to the motion of plasma in a Keplerian accretion disk, immersed in the gravitational well of the compact object, in terms of the inner and the outer radius of the disk, $R_{\rm in}$ and $R_{\rm out}$ (in units of the gravitational radius, $R_g = GM/c^2$, where M is the mass of the compact object), of the index of the assumed power-law dependence of the disk emissivity on the distance from the NS, and of the system inclination, $i$. 
In this way, all the discrete features present in the model are smeared by the same disk parameters. The only difference with respect to the model used by \citet{disalvo_09} is that the \textsc{rdblur} component is now applied also to the iron edge, as it should be if the edge is also produced by reflection in the same disk region as the other emission lines.

As expected, we note a variation in the parameters of the continuum when we exclude the columns from 0 to 7, the most affected parameter being the interstellar absorption column density. 
Its value, and the associated error, progressively increase from $(1.8 \pm 0.1) \times 10^{22}$ cm$^{-2}$ when 0 central columns are excluded (pn-all) to $(3.5 \pm 0.5) \times 10^{22}$ cm$^{-2}$ when 7 central columns are excluded (pn-7). It should be noted here that the best fit value for this parameter obtained with \beppo\ was $(1.9 \pm 0.1) \times 10^{22}$ cm$^{-2}$ \citep{Piraino_07}, compatible  with the results obtained for spectra pn-all to pn-2.
Other significant changes in the parameters values are in the temperature and normalization of the blackbody component, with the temperature decreasing and the normalization increasing from pn-all to pn-7. Also in this case, the best fit values of these parameters obtained with \beppo\ are compatible with the results obtained for spectra pn-all to pn-2.  The temperature of the seed photons of the \textsc{compTT} component is quite stable, while some scattering is observed in the values of the electron temperature and optical depth of this component, although the errors are also very large. Regarding the discrete features and the disk smearing parameters, all the values are compatible with each other within the errors, which of course increase significantly with the decreasing counts.
The inclination angle is the parameter which is mostly affected by the choice of the extraction region, with values ranging 
from $38^{\circ} \pm 1$ to $58^\circ$$^{+23}_{-2}$ going from pn-all to pn-7 (the largest increase occurring starting from spectrum pn-4). 

In order to assess whether the exclusion of some central columns of the CCD has real effects on the spectral shape or simply induce a lack of statistics, 
we carry out the same analysis than described above but considering also the RGS spectrum together with the pn spectrum. The uncertainties on the column density and on the inclination widely decrease. While the value associated to the column density remains constant, $N_\mathrm{H} = (1.6 \pm 0.1) \times 10^{22}$ cm$^{-2}$, the inclination goes from $38^{\circ} \pm 1$ when no central column is excluded to $40^{\circ}$$^{+6}_{-3}$ when 7 central columns are excluded. Also, the values of the parameters such as the blackbody temperature, its normalization and the electron temperature become much more stable, with the error bars considerably reduced. The results are summarized in Table~\ref{tabps:pileup7}.
In conclusion, we checked the effect of pile-up in the EPIC-pn spectrum excluding from 0 up to 7 brightest central columns and comparing the best fit spectral parameters. The parameters seem to depend on the extraction region when looking at the EPIC-pn spectra alone. However, the addition of the RGS data clearly stabilizes the values of the parameters. This means that the variation of the parameters, and of the corresponding error bars, we obtain fitting the EPIC-pn spectra alone are caused by a decrease of the statistics when excluding the central columns of the CCD, and hence by a lack of constraints on the parameters.

Moreover, if we look at the plots presented in Appendix, we can see that the deviation of single and double events at the iron line energy has a minimum in correspondence of the pn-2 spectrum, and it increases again when we exclude more than 2 brightest central columns. This may be ascribed to a mismodeling of the instrumental response and, in particular, of the rate of double events (which involve more than one pixel, and probably more than one column) when too many central columns are excluded. Although most continuum parameters obtained for spectra pn-all to pn-2 are perfectly compatible to those previously obtained with \beppo\ \citep{Piraino_07}, we adopt a conservative approach for the present analysis, to minimize any residual pile-up source of uncertainty. Hence we choose to work with spectrum pn-2.


We therefore considered the spectrum pn-2, whose continuum emission is fitted by a blackbody and a comptonized component (\textsc{compTT}), both modified at low energy by photoelectric absorption.
We first added 4 gaussians and an edge to the continuum model to take into account the emission lines at low energy (S, Ca, Ar), the iron line and the iron edge at 8.6 keV. In this case, the iron line was found at 6.70 ($\pm$ 0.02) keV and its equivalent width was 42 eV. The corresponding $ \chi^{2} _\mathrm{red}$  was 1.14 (408). 

We also tried to fit the iron line at 6.7 keV, likely corresponding to the K$\alpha$ transition of Fe XXV, with the single components of the triplet (that is composed of the forbidden line at 6.64 keV, the intercombination lines at 6.67 keV and 6.68 keV, and the resonance line at 6.70 keV). We also added a line at 6.95 keV associated with emission from Fe XXVI.
To do so, we used 5 gaussians (instead of a single gaussian at 6.6 keV) whose centroid energies are fixed at the expected rest-frame energies.
The lines at 6.67 keV and 6.68 keV are indistinguishable with \xmm,\ since the difference
in the centroid energies is comparable with the fitting errors on the line energy. We therefore considered only one gaussian at 6.67 keV to represent the intercombination lines. This line clearly dominates the triplet, although the lines at 6.64, 6.70 and 6.95 keV contribute slightly to the emission.
The results are summarized in Table~\ref{tabps:triplet}.
Accordingly, we used a single gaussian to describe the Fe complex.

\begin{table}
\begin{minipage}[h]{\columnwidth}
\caption{Iron line complex fitted by 4 gaussians corresponding to the Fe XXV triplet and the Fe XXVI line in the XMM/EPIC-pn data. The energy and the $\sigma$ associated with each transition are in keV.}
\label{tabps:triplet}
\centering
\renewcommand{\footnoterule}{}  
\begin{tabular}{lcccc} 
\hline \hline

\\

\textbf{Fe line}& \textbf{Transition} & \textbf{Energy} & \textbf{$\sigma$} & \textbf{Norm ($10^{-3}$)} \\

   \hline

Fe XXV & $\mathrm{F}$ & 6.64 & $0.11_{-0.11}^{+0.05}$ & $0.6_{-0.5}^{+0.4}$
\\
  & $\mathrm{I_{1}}$ & 6.67 & $0.36 \pm 0.06$ & $2 \pm 0.5$
\\
& $\mathrm{R}$ & 6.70 & $0.09_{-0.09}^{+0.26}$ & $ 0.2_{-0.2}^{+0.6}$
 \\ 
Fe XXVI & $\mathrm{Ly \alpha_{2}}$ & 6.95 & $0.14_{-0.08}^{+0.12}$ & $0.3 \pm 0.2$
\\
\hline
\end{tabular}
\end{minipage}
\end{table}

Then we added a smearing component to this model to take into account relativistic and/or Doppler effects close to the compact object, in the hypothesis of a disk origin of the iron line. This component was convolved with all the 4 gaussians (associated with S, Ar, Ca and Fe) and with the edge. We froze the sigma of the gaussians to zero and apply the same relativistic relativistic smearing parameters to all these features. The iron line is found at  $6.66^{+0.01}_{-0.02}$ keV. The equivalent width is 57 eV. The inclination and the inner radius are  $39^{\circ}$$^{+2}_{-1}$ and $14^{+3}_{-2}$ $R_\mathrm{g}$ respectively. The corresponding $ \chi^{2} _\mathrm{red}$ is 1.03 (409 degrees of freedom, hereafter dof), and the $\chi^2$ decreases by $\Delta \chi^2 = 44$ for one parameter more with respect to the model without the smearing component. Therefore the addition of a mildly relativistic smearing improves the fit, favoring the interpretation of the discrete features in the pn spectrum of \mxb\ as produced by reflection in the inner accretion disk. To give support to this interpretation, in the next  section we fit the broad band ($0.3-200$ keV) spectrum of \mxb\ in the soft and in the hard state to self-consistent reflection models. 

\subsection{Reflection models on pn-2}

In order to test if the iron line, the edge, and the 3 low-energy emission lines (S XVI, Ar XVIII and Ca XIX) are consistent with a reflection scenario, we apply the self-consistent reflection model \textsc{reflionx} on the \xmm/pn spectrum. \textsc{Reflionx} is a self-consistent model including both the reflection continuum and the corresponding discrete features \citep{Ross_05}. In addition to fully-ionized species, the following ions are included in the model: C III-VI, N III-VII, O III-VIII, Ne III-X, Mg III-XII, Si IV-XIV, S IV-XVI, and Fe VI-XXVI. However it does not include Ar XVIII and Ca XIX. We therefore add two gaussians to take into account the emission lines from these two elements. We multiply \textsc{reflionx} by a high energy cutoff (modeled with \textsc{highecut} in xspec), where the cutoff energy is frozen to 0.1 keV and the folding energy is fixed to be 2.7 times the temperature of the electrons. In fact for a saturated Comptonization a Wien bump is formed at the electron temperature, whose peak is at about 3 times the electron temperature.
We use \textsc{nthcomp} \citep{Zdziarski_96, Zycki_99} to model the Comptonization continuum instead of \textsc{compTT} in order to have the photon index $\Gamma$ as fitting parameter of the continuum. Its value is fixed to be equal to the photon index of the illuminating component in the reflection model. 
The $ \chi^{2}$/dof obtained with this model is 577/413 ($\sim 1.40$).

To take into account the smearing of the reflection component induced by Doppler and relativistic effects in the inner disk close to the compact object, we convolve the reflection model and the two gaussians with the same \textsc{rdblur} component. This gives as best fit parameters the inclination of the system ($38^\circ$$^{+2}_{-1}$), the inner and outer radii of the accretion disk ($R_\mathrm{in} = 13^{+4}_{-6}$ and $R_\mathrm{out}$ is fixed at 3500 $R_\mathrm{g}$), and the index of the emissivity law profile (-2.2 $\pm$ 0.1). We freeze the width of the Gaussian lines at 0 keV to apply the same smearing parameters applied to the \textsc{reflionx} component. The $ \chi^{2}$/dof is 523/412  ($\Delta \chi^2 \sim 54$ for the addition of 1 parameter; the corresponding F-test probability of chance improvement is $2 \times 10^{-10}$). 
This attests that the \textsc{rdblur} component is statistically required to improve the fit.
However, the reflection component does not seem to fit correctly the iron edge, since some features are still present in the residuals at about 8.5 keV. We add an edge to the model, also convolved with the same \textsc{rdblur} component  (under the assumption that the edge is also produced in the same region of the accretion disk as the reflection component). The edge is found at 8.7 keV and  the associated depth is 0.06. The $\chi^{2}$ decreases by 83 for the addition of two parameters, resulting in a $ \chi^{2} _\mathrm{red} = 1.07$ (410). 

To check the consistency of the edge with the reflection continuum, we used another reflection model, \textsc{pexriv} \citep{magdziarz_95}, which includes the iron edge and the Compton bump of the reflection component. This model consists in an exponentially cutoff power-law spectrum reflected by ionized material. However this model does not include any emission lines, so 4 gaussians (S, Ar, Ca and Fe) were added and convolved, together with the pexriv component, with the \textsc{rdblur} component. 
The photon index of \textsc{pexriv} is fixed to the one obtained with the \textsc{nthComp} model and the cutoff energy is 2.7 times the electron temperature obtained with \textsc{nthComp}.
The normalization of the \textsc{pexriv} model has been fixed to the one of the cutoff power-law model included in \textsc{nthComp}. To do so, we applied a cutoff power-law model to the data and calculated the normalization in such way that the bolometric flux is the same than with \textsc{nthComp}.
The $ \chi^{2} _\mathrm{red}$ obtained with this model is 1.06 (410 dof). The values of the smearing parameters are in agreement with the previous model and the features at about 8.5 keV are not visible anymore in the residuals. So the edge is well fitted by \textsc{pexriv} and is likely a reflection feature. However, it is not well fitted by the \textsc{reflionx} component. 

\subsection{Reflection models on the \xmm,\ \beppo\ and \rxte\ data}

\begin{figure}[t]
\includegraphics[height=6.0cm]{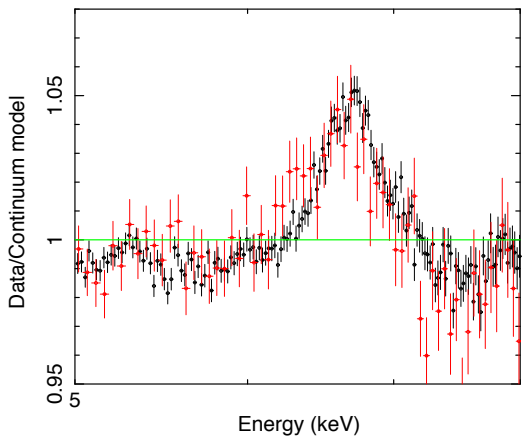}
\caption{Ratio of the data to the continuum model for XMM/EPIC-pn (black) and BeppoSAX/MECS (red) in the energy range $5-8$ keV showing the broad iron line clearly visible in both spectra, although the statistics is much better in the case of the pn spectrum.
The continuum is described by the spectral model phabs (bbody + compTT) in Xspec.}
\label{fig:ratio_SS}
\end{figure}

Considering separately the data of \xmm\ and \beppo,\ the continuum of these spectra is 
well fitted by a blackbody plus a thermal Comptonized model (\textsc{compTT} or \textsc{nthcomp}) modified at low energy by the interstellar photoelectric absorption (\textsc{phabs}). 
In both cases, the fit is improved by adding a broad iron line, fitted by a Gaussian line or, even better, by a \textsc{diskline}. 
The best fit parameters, obtained fitting separately the spectra from the two different X-ray observatories, are  similar to each other and are in agreement with previous results reported by \citet{Piraino_07} and  \citet{disalvo_09} on \mxb.\ Regarding the \beppo\ data, the absorption column density is $N_\mathrm{H} = 1.4 \times 10^{22}$ cm$^{-2}$, the blackbody temperature is 0.56 keV, the temperatures of the electrons and of the seed photons of the Comptonized component are 3.5 keV and 1.2 keV, respectively. A diskline is found at 6.8 keV, the inclination of the system is $28^\circ$$^{+8}_{-5}$  and the inner radius is about $8^{+4}_{-2}$ $R_\mathrm{g}$. The $ \chi^{2}$/dof corresponding to this fit is 584/503 ($\chi^2_{red} \sim 1.16$). This is better than using a Gaussian to fit the iron line ($ \chi^{2}$/dof is 618/505, $\chi^2_{red} \sim 1.23$; an F-test gives a probability of chance improvement of about $10^{-6}$). These results are in perfect agreement with the results we obtain from the \xmm\ spectrum, independently of the particular model we use to fit the reflection features.
Note, however, that the uncertainties on the inner radius and on the inclination angle are larger than in the case of \xmm.\ This can be explained by the quality of the data, which is better in the case of \xmm\ due to its larger effective area and higher resolution capabilities. In Fig.~\ref{fig:ratio_SS} we show the ratio of the data to the best-fit continuum model in the energy range 5$-$8 keV to compare the residuals at the iron line as observed by the XMM-Newton/EPIC-pn and by the BeppoSAX/MECS. The iron line profile appears very similar in the two instruments, although the observations are not simultaneous.

Since the values of the parameters obtained by fitting the \xmm\ and \beppo\ spectra in the soft state are very similar, we fitted these data simultaneously, adding also the \rxte\ data. The different cross calibrations of the different instruments were taken into account by including normalizing factors in the model. This factor was fixed to 1 for pn and kept free for the other instruments.
We use the self-consistent reflection model \textsc{reflionx},
and \textsc{nthcomp} instead of \textsc{compTT} to describe the thermal Comptonization component in order to have the photon index as a parameter of the fit. To take into account the disk smearing, which is necessary to obtain a good fit of the reflection component, we convolve the reflection model, the edge and the two Gaussians used to fit the Ar and Ca lines with the same \textsc{rdblur} component. 

Being the spectra not simultaneous, we find small differences in the best fit values of some parameters of the continuum model. We therefore let these parameters free to vary from one instrument to the other, when necessary (see Table~\ref{tabps:SSparam}). These parameters are the column density, the parameters of \textsc{nthcomp} (the photon index, the temperature of the electrons, of the seed photons, and the normalization) and the fold energy of the high energy cutoff (that is fixed at 2.7 times the electron temperature in the soft state, according to the expectation for saturated Comptonization), which are slightly different for \beppo\ and \xmm.\ 
 As regards \rxte,\ only the column density and the photon index are left free to vary. The other parameters coincide very well with those found for the \beppo\ spectrum, so they are constrained to have the same values.
All the other parameters are perfectly consistent with those obtained for \xmm, and were forced to have the same values. The total $ \chi^{2}$/dof obtained in this way is 1839/1573 ($\chi^2_{red} \sim 1.17$).

\begin{table}
\begin{minipage}[h]{\columnwidth}
\caption{Parameters left free to vary between \xmm,\ \beppo,\ and \rxte\ spectra in the soft state. The whole model applied to these data is presented in Table~\ref{tabps:SStot}. The letter "B" indicates that the corresponding parameter has the same value as in the \beppo\ spectrum.} 

\label{tabps:SSparam}
\centering
\renewcommand{\footnoterule}{}  
\begin{tabular}{lccc} 
\hline \hline

\\

\textbf{Parameter}& \textbf{XMM} & \textbf{BeppoSAX} & \textbf{RXTE} \\

   \hline

 $ N_\mathrm{H} ~ (\times 10^{22}cm^{-2})$ & $2.08 \pm 0.02 $ & $1.96 \pm 0.02 $& $3.64 \pm 0.02$
\\
 $ \Gamma $ & $ 2.6 \pm 0.1 $ & $ 2.2 \pm 0.1 $ & $ 2.4 \pm 0.1 $
 \\
 $ kT_\mathrm{e}$ (keV) & $ 3.0 \pm  0.1 $ &  $ 2.9 \pm  0.1 $ & B
 \\ 
 $ kT_\mathrm{seed}$ (keV) & $ 1.30 \pm  0.02  $ & $ 1.13^{+0.01}_{-0.02} $ & B
 \\
$Norm_\mathrm{nthComp}$  & $ 0.14 \pm 0.01 $ & $ 0.19 \pm  0.01 $ & B
\\
\hline
\end{tabular}
\end{minipage}
\end{table}

We note that the \textsc{rdblur} component is not statistically required to fit the whole dataset in this case. Indeed, if we delete this component from the model, we obtain $\chi^2 = 1835/1574$ ($\sim 1.17$), which is very similar to the one we get when the relativistic smearing is included in the model ($ \chi^{2}/dof = 1839/1573$). However, 
if we exclude the relativistic smearing from the model, the ionization parameter gets an extremely high value, $\sim 8000$ erg cm s$^{-1}$, that appears to be unphysical. At such a high ionization parameter, Fe XXVI  would be the most abundant Fe ion and this would produce a line at 6.97 keV. On the other hand, in the \xmm\ spectrum the iron line is clearly detected at 6.7 keV, suggesting it is produced by Fe XXV.
With the inclusion of the relativistic smearing described by the \textsc{rdblur} component, the ionization parameter attains a more reasonable value of 3500, fully consistent with the presence of an iron line at 6.6 keV produced by Fe XXV. Moreover, all the smearing parameters 
are perfectly coherent with those previously obtained \citep[e.g.][]{disalvo_09,dai_10,Piraino_07}.
An additional iron edge is found at 8.7 keV with a significance of 11.6 $\sigma$.
Finally, we report here, for the first time, evidences of an iron overabundance by a factor 2.5 with respect to its solar abundances. The results of this model are presented in Table~\ref{tabps:SStot}.
In order to evaluate the statistical significance of the iron overabundance, we fixed this parameter to 1. The $ \chi^{2}$ increases by 48 for the addition of 1 d.o.f. ($\chi^2_{red} \sim 1.20$) and the associated probability of chance improvement is $2 \times 10^{-10}$. So the iron overabundance is statistically significant.

The EPIC/pn absorbed flux obtained from the \xmm\ best fit spectral parameters is $6.19 \times 10^{-9}$ erg cm$^{-2}$ s$^{-1}$ and the unabsorbed flux is $7.39 \times 10^{-9}$ erg cm$^{-2}$ s$^{-1}$ in the 2$-$10 keV band. We extrapolate this model in the 0.1$-$150 keV range to estimate the bolometric unabsorbed flux, $F_\mathrm{X} = 2.7(1) \times 10^{-8}$ erg cm$^{-2}$ s$^{-1}$. The bolometric luminosity associated to the soft state is $L_\mathrm{X} \sim 1.8 \times 10^{38}$ erg s$^{-1}$ assuming a distance to the source of 7.4 kpc. This value is very close to the Eddington luminosity for a 1.4 $\mathrm{M_{\odot}}$ neutron star.

\begin{figure}[h]
\includegraphics[height=8.9cm,angle=-90]{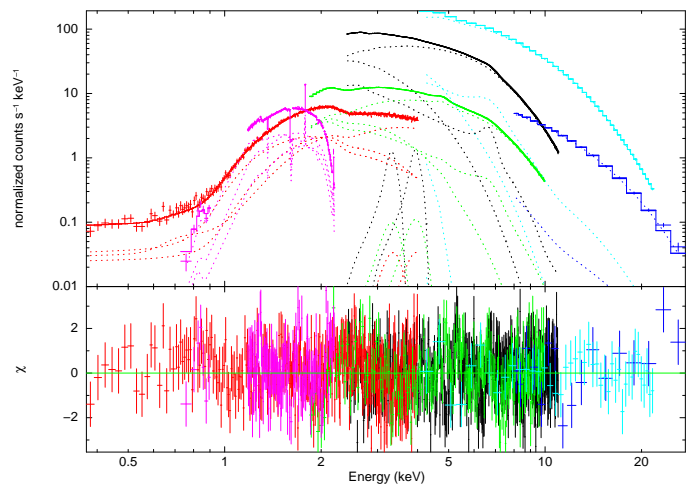}
\caption{Top panel: XMM/RGS1 (magenta), XMM/EPIC-pn (black), BeppoSAX/LECS (red), BeppoSAX/MECS (green), BeppoSAX/HPGSPC (blue), RXTE/PCA (cyan) data points in the range $0.3-30$ keV, when \mxb\ was in the soft state. Bottom panel: Residuals (data-model) in unit of $\sigma$ when the \textsc{reflionx} model is used to describe the reflection component. The parameters associated to this model are indicated in Table~\ref{tabps:SStot}. The XMM/RGS1 data have been rebinned for graphical purpose.}
\label{fig:reflionx}
\end{figure}




We also apply another reflection model, \textsc{xillver} \citep{garcia_10}, to our dataset of \mxb\ from the three satellites. 
This model includes Compton broadening, and the illumination spectrum is a power law with a photon index of 2, similar to \textsc{reflionx}. In this model, the redistribution of the photon energy is achieved by a Gaussian convolution, whose sigma is a function of the energy and the temperature of the gas. The gas temperature changes going deep inside the disk, and is calculated self-consistently by solving thermal and ionization balance.
This model also includes emission lines from the same ions included in \textsc{reflionx} and, in addition, emission lines from Ar and Ca. It also allows to fit the abundances of these two elements with respect to the solar abundance. In this case, to obtain a stable fit, we were forced to freeze the photon index $\Gamma$ associated to the \xmm\ data to the value obtained with \textsc{reflionx} ($\Gamma$ = 2.6).
The addition of the edge at 8.5 keV is again necessary to improve the fit, with a significance of 11.1 $\sigma$. 
The $ \chi^{2}_{red}$ associated to this fit is 1.26 for 1576 dof. 
The fit obtained with this model is a bit worse than that obtained with \textsc{reflionx}, but the values of the parameters are still consistent with those obtained with \textsc{reflionx}. We note a lower value of the inclination angle of the system ($i =  25-27^{\circ}$), while the value of the inner disk radius ($R_\mathrm{in} = 10-13$ $R_\mathrm{g}$) is perfectly consistent between the two cases\footnote{We use here a test version of the \textsc{xillver} model, specifically developed to include Ar and Ca lines. In this version of the model, the illuminating flux is integrated over a different energy range with respect to that used for the \textsc{reflionx} model, and this results in a different, lower, value of the ionization parameter, which also affects the estimate of the inclination angle (J. Garcia, private communication). We have checked that, integrating the illumination flux over the same range used for the \textsc{reflionx} model, we obtain best-fit parameters that are all compatible with those obtained with \textsc{reflionx}.}. We find an overabundance by a factor 1.5$-$2 of Ar and Ca with respect to their solar abundance; a similar overabundance is also observed for iron (see Table~\ref{tabps:SStot}).  However the statistical significance of the overabundance of Ar and Ca is not very well established. Indeed, fixing this parameter to 1, we find $\chi^{2}/dof = 1992/1577$, the corresponding probability of chance impovement is $2 \times 10^{-2}$. In any case, this model allows to demonstrate that the Ar and Ca lines are likely produced by reflection.

As seen in the previous section, \textsc{pexriv} also gives a very good fit of the pn data.
We apply this model plus 4 gaussians, all convolved with the same rdblur component, to the whole dataset.
The results are very similar to \textsc{reflionx}, in particular the inclination of the system ($i =  38-41^{\circ}$) and the inner radius of the accretion disk ($R_\mathrm{in} = 12-17$ $R_\mathrm{g}$). The $\chi^{2}_{red}$ associated is 1.16 for 1570 dof.
The best fit parameters corresponding to this model are summarized in Table~\ref{tabps:SStot}
and compared to \textsc{reflionx} and \textsc{xillver}. 

\section{Hard state}

In this section we apply the same procedure used for the soft state to the \mxb\ data in the hard state, using the three satellites: \xmm,\ \beppo\ and \rxte.\
Regarding the \xmm\ data, we excluded a type-I X-ray burst before performing the spectral analysis. The study of the burst is described by \citet{dai_10}. 

\begin{figure}[t]
\includegraphics[height=6.0cm]{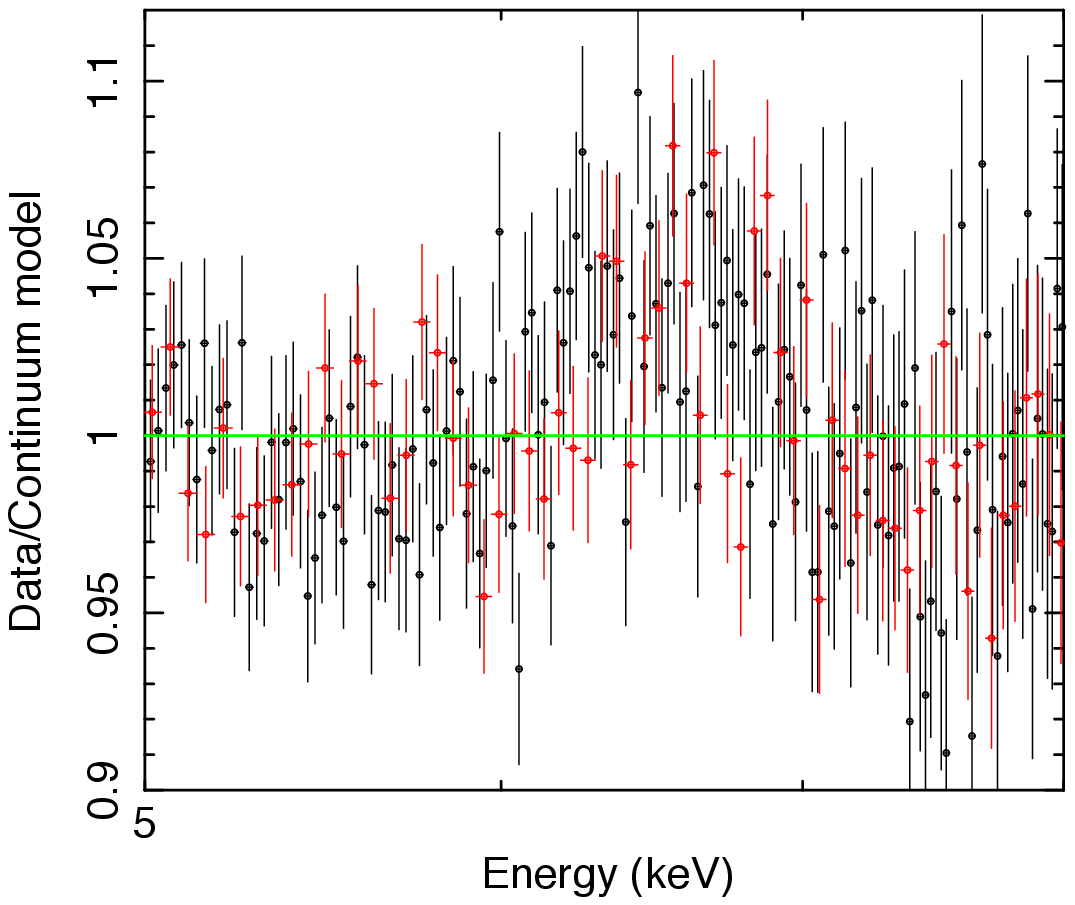}
\caption{Ratio of the data to the best-fit continuum model for XMM/EPIC-pn (black) and BeppoSAX/MECS (red) in the hard state, in the energy range $5-8$ keV. The broad iron line is visible in both spectra, but is not as well defined as in the soft state (see Fig.~\ref{fig:ratio_SS}). The continuum is described by phabs (bbody + compTT).}
\label{fig:HS_feline}
\end{figure}

We use the same continuum model in order to determine the differences in the spectral parameters from one state to the other.
The temperature of the electrons is about 14$-$16 keV, the temperature of the seed photons is 1.0$-$1.2 keV, the optical depth of the Comptonized component is 5$-$6 and the blackbody is found at 0.55$-$0.58 keV. The addition of a gaussian improves considerably the fit. A broad iron line is present in all the data at 6.4$-$6.6 keV (see Fig.~\ref{fig:HS_feline}). 
Using a \textsc{diskline} instead of a Gaussian profile to fit the iron line, we have to freeze the values of the outer radius to 
3500 $R_\mathrm{g}$ and the inclination of the system to 37$^{\circ}$, 
as in the soft state and in agreement with \citet{disalvo_09}. The $ \chi^{2}$ in this case is the same than using a simple Gaussian line.
While for the \xmm\ data, the $ \chi^{2}$/dof is 407/419  ($\sim 0.97$) using a gaussian, there are still some residuals at high energy ($> 10$ keV) in the case of \beppo\ and \rxte.\ The associated $ \chi^{2}_\mathrm{red}$ are 1.24 (484 dof) and 1.15 (65 dof) for the \beppo\ and the \rxte\ spectra, respectively.

In order to fit this hard excess, probably ascribed to the Compton bump of the reflection component, we apply the self-consistent reflection model \textsc{reflionx}, with the aim of comparing the values of the parameters in the hard state and in the soft state. In this case we fix the folding energy of the high energy cutoff  to the electron temperature of the \textsc{nthcomp} component, as expected for non saturated Comptonization.
We simultaneously fitted the three datasets following the same method described in previous section.
Some parameters are left free to vary between the instruments of the different observatories such as the column density and the blackbody temperature. 
Moreover, in the case of the \rxte\ spectrum, we had to let the photon index of the power-law and the electron temperature free to vary in order to obtain a good fit. 
These parameters are listed in Table~\ref{tabps:paramHS}.
The inclination angle is not well constrained so we freeze its value to 37$^{\circ}$, 
which corresponds to the best fit value obtained in the soft state using \textsc{reflionx} to fit the reflection component. The outer radius of the accretion disk is also frozen to 3500 $R_\mathrm{g}$. This value corresponds to the best estimate obtained using the \textsc{reflionx} model to fit the soft state of \mxb\ and is consistent with the results reported by \citet{disalvo_09}.

\begin{table}
\begin{minipage}[t]{\columnwidth}
\caption{Parameters left free to vary between \xmm,\ \beppo,\ and \rxte\ in the hard state. The complete model is given in Table~\ref{tabps:HStot}. $ kT_\mathrm{bb}$ and $Norm_\mathrm{bb}$ refer to the \textsc{blackbody} component whereas $ \Gamma $ and $ kT_\mathrm{e}$ concern the \textsc{nthComp} model. The letter "B" means that the value of the corresponding parameter is fixed to the one of the \beppo\ spectrum.}

\label{tabps:paramHS}
\centering
\renewcommand{\footnoterule}{}  
\begin{tabular}{lccc} 
\hline \hline

\\

\textbf{Parameter}& \textbf{XMM} & \textbf{BeppoSAX} & \textbf{RXTE} \\

   \hline

 $ N_\mathrm{H} ~ (\times 10^{22}cm^{-2})$ & $2.0 \pm 0.3$ & $1.9 \pm 0.1$ & B
\\
 $ kT_\mathrm{bb}$ (keV) & $ 0.58^{+0.04}_{-0.02} $ & $ 0.24^{+0.02}_{-0.03} $ & B
\\
$Norm_\mathrm{bb}$ $(\times 10^{-3}) $ & $ 2.6 \pm 0.5 $ & $ 2.0 \pm 0.4 $  & B 
 \\
 $ \Gamma $ & B & $ 1.84 \pm 0.01 $ & $ 2.07^{+0.05}_{-0.03} $
 \\
 $ kT_\mathrm{e}$ (keV) & B & $ 22^{+2}_{-1} $ & $ 79^{+50}_{-22} $
 \\ 
\hline
\end{tabular}
\end{minipage}
\end{table}

We convolve the reflection model with the \textsc{rdblur} component to include Doppler broadening
caused by the motion of matter in the inner disk. The addition of this component decreases the $\chi^2 /dof$ from $1126 /981$ to $1097 / 980$.
The F-test is $\sim 4 \times 10^{-7}$. This demonstrates that the iron line and the whole reflection component are significantly broadened also in the hard state, and are consistent with being produced in the accretion disk.
Since only the iron line is significantly detected, we did not add any other emission line, nor edge.
To measure the significance of the Compton bump in these data, we replace the \textsc{reflionx} model with a simple gaussian line smeared by relativistic and/or Doppler effects, by using
the \textsc{rdblur} component. The $\chi^2$ is $1299/981 \sim 1.32$. We compare this result with the previous model where the reflection component is included instead of the simple gaussian associated to the iron line. The F-test gives a probability of chance improvement equals to $10^{-38}$. This shows that there is a reflection signature in the spectrum beside the iron line in the hard state, meaning the Compton bump is significantly detected in these data.

\begin{figure}[h]
\includegraphics[height=8.9cm,angle=-90]{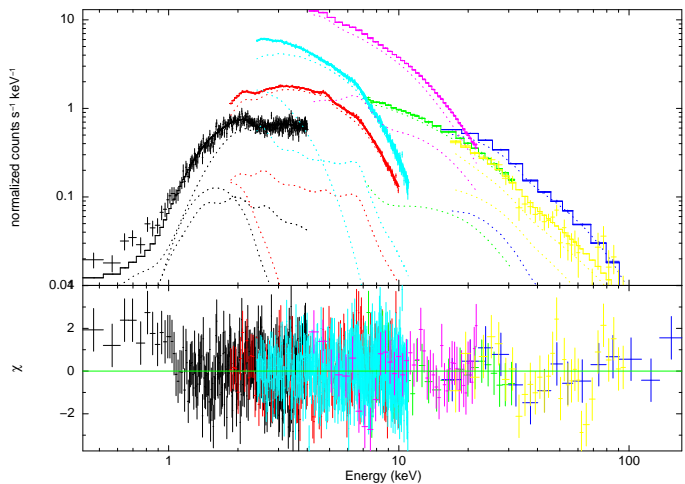}
\caption{Top panel: XMM/EPIC-pn (cyan), BeppoSAX/LECS (black), BeppoSAX/MECS (red), BeppoSAX/HPGSPC (green), BeppoSAX/PDS (blue), RXTE/PCA (magenta), RXTE/HXTE (yellow) data points in the range $0.4-200$ keV, corresponding to the hard state of \mxb.\ Bottom panel: Residuals (data-model) in unit of $\sigma$ when the \textsc{reflionx} model is applied. The parameters associated to this model are shown in Table~\ref{tabps:HStot}.}
\label{fig:reflionx}
\end{figure}


We obtain a good fit of the whole dataset;
the best fit parameters are listed in Table~\ref{tabps:HStot}, where we just show those obtained for the \beppo\ spectrum (there exist small differences 
in the best-fit parameters related to the fact that the observations are not simultaneous as described above).
From all the instruments, we obtain a constraint on the inner disk radius, which appears to be truncated further from the neutron star surface ($R_\mathrm{in} = 
19-59$ $R_\mathrm{g}$) with respect to the soft state.
We obtain a high temperature of the electrons (about 22 keV) in the hard state, 
and a low value of the ionization parameter ($\xi \sim 210$ erg cm s$^{-1}$). 

The absorbed flux associated to this model obtained from the \beppo\ best fit parameters in the range 0.1$-$150 keV is $ 2.09 \times 10^{-9}$ erg cm$^{-2}$ s$^{-1}$, and the unabsorbed flux is $ 2.99 \times 10^{-9}$ erg cm$^{-2}$ s$^{-1}$. The bolometric luminosity associated to the hard state is $L_\mathrm{X} \sim 2 \times 10^{37}d^{2}_{7.4}$ erg s$^{-1}$
which corresponds to $\sim  11\%$ $ L_\mathrm{Edd}$.

Because in the soft state, an iron overabundance has been detected by at least a factor 2, we fixed this parameter to 2 in order to evaluate the change in the whole spectrum and its effect on the other parameters. The values of the parameters are very similar, even if the $\chi^2$ is a bit higher (1.13 for 981 dof). The inner disk radius is found at a slightly larger distance from the compact object ($R_\mathrm{in} = 24-79$ $R_\mathrm{g}$), again in agreement with a geometry where the accretion disk is truncated further from the neutron star than in the soft state. These results are summarized in Table~\ref{tabps:HStot}.


\section{Discussion}

We performed a broad band and high energy resolution spectral analysis of \mxb\ using data from \xmm,\ \beppo\ and \rxte, both in the hard and in the soft state. We detailed a method to fit non simultaneous data, that we applied using self-consistent reflection models. Due to the good spectral coverage from the low (0.3 keV) to the high (200 keV) energy band, we have obtained good constraints on the continuum emission and on the reflection component, 
marking the most significant changes in the spectral parameters
from the hard state to the soft state. The agreement between the results obtained with different instruments and the possibility to fit most of the reflection component with self-consistent models, provides further evidence that the broad iron line observed in \mxb\ may be produced by reflection at the inner accretion disk, and indicates that the reported inner disk parameters are indeed reliable. In the following we discuss the main results we have obtained from this analysis.

\subsection{The iron line}

The origin of the iron line in neutron star LMXBs is still debated \citep{Ng_10, Cackett_10, Cackett_12}.
In this paper, we have compared the iron line profile observed by \xmm\ with that observed by other instruments, and in particular by \beppo. We find a perfect agreement between the results from these instruments, although the observations were not simultaneous, both in the soft and in the hard state of the source. 

We studied carefully the pile-up effects in the XMM/EPIC-pn spectrum of \mxb\ during the soft state, when the count rate was the highest. We have demonstrated that the exclusion of the 2 central columns of the CCD is enough to get rid of pile-up effects, while excluding more CCD rows may underestimate the number of double events with respect to the expected value (see Appendix). In addition to this, we showed that the addition of the RGS data is important to 
well constrain the overall continuum shape.
We find that the iron line parameters are not significantly affected by pile-up. This is in agreement with \citet{Cackett_10}, who also conclude that the iron line profile is robust even if there was a small fraction of pile-up affecting the data, while the continuum may vary significantly. This is also in agreement with a comprehensive study performed by  \citet{miller_10} who found that severe pile-up may distort 
disk lines and the continuum shape whereas a modest pile-up fraction does not sensibly affect the line shape. This is probably the case of \mxb.\
The best fit parameters of the iron line profile we find in this way are perfectly in agreement with what has been previously found \citep[see e.g.][]{Piraino_07, disalvo_09, dai_10}, independently from the particular model used to fit the reflection component.

In the soft state, the Fe line at 6.7 keV is associated with highly ionized Fe XXV, which is a triplet consisting of the following components: at $r = 6.700$ keV, \,$i_{2} = 6.682$ keV, $i_{1} = 6.668$ keV, and $f = 6.637$ keV.  We also included a gaussian to consider the H-like Fe XXVI contribution of the Ly$\alpha$ transitions at $Ly\alpha_{1} = 6.973$ keV and $Ly\alpha_{2} = 6.952$ keV. Unfortunately, the resolution of \xmm\ and \beppo\ does not allow us to resolve the structure of the resulting line, which appear to be dominated by the intercombination line of the triplet, and this is why a single Gaussian or a diskline has been used to take into account all these components. 
These lines were resolved, for example, in the case of the bright Z-source Cyg X-2 observed with the High Energy Transmission Grating Spectrometer onboard of the \chandra\ satellite, that offers very high spectral resolution \citep{Schulz_09}. Also in that case the Fe complex was dominated by the intercombination line of the Fe XXV triplet and therefore the line was fitted by a single Gaussian. Note also that a red-skewed wing of the iron line was discovered in a \suzaku\ observation of Cyg X-2 \citep{Shaposhnikov_09}.
It should be noted, however, that these lines are all included in the reflection models we used in our spectral analysis, and a further smearing was required to properly fit the line complex. Consequently the iron line is consistent with being produced in the inner part of the accretion disk where the line profile is distorted by Doppler and mildly relativistic effects relatively close to the compact object. At the inner disk radii we find, $R_{\rm in} \sim 10-17\, R_g$, the Keplerian velocities become mildly relativistic and the Doppler boosting effect yields the blue-shifted horn (produced by matter coming in our direction) brighter than the red-shifted one (produced by receding matter).

In the hard state, the Fe emission line at $6.4-6.6$ keV is related to a low ionized Fe fluorescence line.
 The line does not present a clear asymmetry anymore and is equally well fitted by a gaussian or with the diskline model. 
The apparent symmetry of the line may be due to the relativistic effects becoming less important further from the compact object, and/or to the lower statistics in the hard state.
In both states, the broadening of the line is not as extreme as in the case of some black hole X-ray binaries or AGNs \citep[see e.g.][]{Reis_09b, Fabian_09}.
The Compton hump of the reflection component, however, is required with a very high confidence level to get a good fit of the data in the hard state. 



\subsection{Reflection models}

We used data of \mxb\ from three satellites (\xmm,\ \beppo,\ and \rxte)\ in order to test self-consistent reflection models on a broad-band range, from 0.3 to 200 keV.
The reflection models we used are calculated for an optically-thick atmosphere (such as the surface of an accretion disk) of constant density illuminated by a power-law spectrum. 

In the soft state, we conclude that the reflection model \textsc{pexriv} is able to fit the iron edge with smearing parameters very similar to those obtained for the iron line profile; the \textsc{reflionx} model is able to fit self-consistently the iron line profile and the S XVI line at 2.6 keV; the \textsc{xillver} model is able to fit the iron line profile self-consistently with the S XVI, the Ar XVII, and Ca XIX lines. 
The \textsc{reflionx} model suggests an iron overabundance by a factor of about 2
with respect to its solar abundance. 
A mildly relativistic smearing of the reflection component, described with the \textsc{rdblur} component, is statistically required to fit the \xmm\ data with the \textsc{reflionx} model, these data having the best statistics at the iron line energy. Nevertheless, this component is not statistically required when we fit all our dataset with the \textsc{reflionx} model.
However, if we exclude the \textsc{rdblur} component from the model, we find that the ionization attains very high (unphysical) values. It should be noted here that both \textsc{reflionx} and \textsc{xillver} include compton broadening, that is higher for higher ionization parameters \citep{Reis_09}. This means that to adequately fit the width of the line with Compton broadening only, we have to increase dramatically the ionization parameter to values which appear too high to be compatible with the observed energy of the iron line. This is why we decided to include the relativistic smearing in the reflection model.
The smearing parameters are well determined and appear very similar regardless of the particular reflection model adopted. They are congruent with previous results reported for this source \citep[][]{Piraino_07, disalvo_09, dai_10}, and are very similar for all the datasets used here (from \xmm,\ \beppo,\ and \rxte\ satellites). 

In the hard state, we significantly detect the iron line at 6.4 keV and the Compton bump of the reflection component, while the other low-energy emission lines and the iron edge were not significantly detected. We just used the \textsc{reflionx} model to fit self-consistently the reflection component (i.e.\ just the iron line and the Compton bump in this case). As before, we obtain a very good fit of the whole spectrum in the broad energy band between 0.4 and 200 keV.
We note that the relativistic smearing is statistically required for the hard state. In fact the iron line is found to be broad also in this state (although not as broad as in the soft state), but in that case the ionization parameter has a particularly small value, Compton broadening is negligible and cannot explain the width of the line. Our results for the hard state of 4U 1705-44 are  coinciding with the results obtained by Piraino et al. (in preparation); these authors fit the \beppo\ data in the hard state with an alternative reflection model, where \textsc{compPS} \citep{Poutanen_96} is used to model the primary Comptonization spectrum. Their results are consistent with the disk-reflection scenario we favor in this paper.

Our results are in agreement with \citet{Reis_09} who applied \textsc{reflionx} on three datasets of \mxb\ obtained by \suzaku\ at different periods. A relativistic broadening was also required to obtain a good fit, as in our analysis. Therefore different instruments get similar values for the reflection component in \mxb,\ indicating that the iron line shape as seen by these instruments 
is similar, and demonstrating that pile-up is not responsible of the observed iron line shape.

\citet{dai_10} and \citet{Cackett_10} also applied a reflection model, \textsc{refbb} \citep{Ballantyne_04}, to fit the spectrum of this source, restricted to the \xmm\ data in the energy range 2--12 keV. In this model a blackbody component, likely associated to the boundary layer, provides the illuminating flux. In this energy band, the reflection model parameters are mainly determined by the shape of the iron line, and are perfectly compatible with the results described in this work.
In other words, the inner disk parameters we obtain for this source under the hypothesis that the line is produced by reflection of the primary Comptonization spectrum on the inner accretion disk, are always compatible with each other, independently of the particular model used to fit the reflection component (such as \textsc{disklines}, \textsc{refbb}, \textsc{pexriv}, \textsc{reflionx}, \textsc{xillver}). 

To support this disk-reflection scenario, all the reflection component should be fitted by a 
self-consistent reflection model.
Unfortunately no reflection model includes all the reflection features. In the case of \textsc{reflionx}, two emission lines are necessary to reproduce Ar XVIII and Ca XIX emission lines detected in the soft state. We also needed to add an edge at $\sim 8.5$ keV. In order to check whether the edge was consistent with a disk origin, we used another reflection model: \textsc{pexriv}. The edge is well taken into account by this model, however this model does not include any emission line. The \textsc{xillver} model includes S, Ar, Ca and Fe lines
but the addition of an edge at $\sim 8.5$ keV was again necessary to obtain a good fit.

As shown by our analysis, the quality reached by today's observatories is such that it is now compelling to better calibrate the reflection broad band spectrum, as important features are clearly present (like emission lines from low Z-elements) and others are not well accounted for (as absorption edges of highly ionized elements).

\subsection{The continuum parameters}

The main differences between the spectral parameters obtained from the soft and the hard state are discussed in the following. They concern the electron temperature and
the seed photon temperature of the Comptonized component,
the inner radius of the disk as derived from the smearing of the reflection component,
and the ionization of the reflection component.
We note here that we had to fix the parameters of the continuum of the reflection component to the parameters of the \textsc{nthcomp} component used to represent the primary Comptonization spectrum. In this case, in order to obtain a good fit, we had to fix the folding energy parameter to 2.7 times the electron temperature in the soft state and to the electron temperature in the hard state, in agreement with the fact that the peak of a saturated Comptonization spectrum is at 2.7 $kT_e$, while the peak of a non saturated Comptonization spectrum is at $kT_e$.

We observe a clear difference in the spectral parameters from the soft to the hard state. The electron temperature increases (from $kT_\mathrm{e} =2-3 $ keV in the soft state to $\sim 20-24$ keV in the hard state) whereas 
the power-law photon index and the temperature of the seed photons decrease (from $ \Gamma = 2.2-2.8 $ to $ \Gamma = 1.8$ and from $kT_\mathrm{seed} = 1.1-1.4 $ keV to $0.7-0.8$ keV, respectively). 

In order to evaluate the changes in the optical depth and the region of the seed photons
from one state to the other, we use the parameters obtained by the \textsc{nthComp} model. This model specifies the Comptonization via the electron temperature in the corona $kT_\mathrm{e}$, the temperature of photons injected in the corona $kT_\mathrm{seed}$, and the spectral slope $\Gamma$. These parameters are related to the optical depth $\tau$ as
\begin{equation}
\Gamma = -\frac{1}{2} + 
\left[ \frac{9}{4} + \frac{1}
{\frac{kT_\mathrm{e}}
{m_\mathrm{e}c^{2}} \, \tau \, (1 + \frac{\tau}{3})
}
\right]^{1/2}
\end{equation}
\citep{Lightman_87}. We use the values of the \textsc{nthComp} components reported in the first column of Tables~\ref{tabps:SStot} and Table~\ref{tabps:HStot}, which correspond to the \textsc{reflionx} model for the reflection component in both states. 
The optical depth decreases from the soft to the hard state, from $\tau \sim 7$ to $\tau \sim 3$.
Moreover we compute the Comptonization parameter $y$ defined by
\begin{equation}
y = 4 \frac{kT_\mathrm{e}}{m_\mathrm{e} c^{2}} \times \max{(\tau, \tau^{2})}
\end{equation}
and we obtain $y \sim 1$ and $y \sim 2$ in the soft and hard state respectively. 
To estimate the radius of the emitting region of the seed photons which are Comptonized in the hot corona, we assume their emission as a blackbody and the bolometric Comptonized flux as $F_\mathrm{Compt} = F_\mathrm{seed} \, (1 + y)$ since we have to correct for energy gained by the photons in the inverse Compton scattering. Hence, $F_\mathrm{seed}$ is defined by
\begin{equation}
F_\mathrm{seed} = \sigma T^{4}_\mathrm{seed} \left( \frac{R_\mathrm{seed}}{d} \right)^{2}
\end{equation}
and so the region of the seed photons is obtained by
\begin{equation}
R_\mathrm{seed} = 3 \times 10^{4} d 
\frac{[F_\mathrm{Compt} / (1 + y)]^{1/2}}
{(kT_\mathrm{seed})^{2}} \, \textrm{km}
\end{equation}
\citep[see][]{intzand_99} with $d$ the distance in kpc, $F_\mathrm{Compt}$ in $\textrm{erg cm}^{-2} \, \textrm{s}^{-1}$, and $kT_\mathrm{seed}$ in keV. By considering the $F_\mathrm{Compt}$ obtained with \beppo\ that is $8.84 \times 10^{-9}$ and $2.09 \times 10^{-9}$ $\textrm{erg cm}^{-2} \, \textrm{s}^{-1}$ in the soft and hard state, respectively, and a distance of 7.4 kpc, we obtain $R_\mathrm{seed} \sim 9 $
km in the soft state and $R_\mathrm{seed} \sim 12 $
km in the hard state.
So the seed photons are compatible with coming from the neutron star surface in both states. 
These results are consistent with a corona above the disk and/or between the disk and the stellar surface which is hotter in the hard state than in the soft state. This is likely due to the interactions with the soft photons from the disk that are more or less intense depending upon the geometry of the disk-corona system.
It can also be connected with the energetic balance between the Compton cooling provided by the soft photons (which acts as the photon number, so with the fourth power of the temperature) and the coronal heating (maybe through shock dissipation).

For what concerns the blackbody component, in the soft state $kT_\mathrm{bb}$ is 0.57 keV. Assuming a distance of 7.4 kpc, the region associated with the blackbody has an apparent radius $\sim R_\mathrm{bb} = 30$ km, in agreement with the emission coming from the hottest part of the accretion disk, which corresponds to the inner part of the disk close to the neutron star. 
In the hard state, the blackbody temperature is  $kT_\mathrm{bb} = 0.25$ keV for all the instruments except for \xmm\ for which $kT_\mathrm{bb} = 0.57$ keV. The radius associated to the first value is $\sim R_\mathrm{bb} = 160$ km, again compatible with a truncated accretion disk  
 whereas the second one is $\sim R_\mathrm{bb} = 30$ km.
It is not clear whether the temperature of the blackbody component changes between the soft and the hard state, since for this parameter we find a value for the hard state very similar to what is found for the soft state for the \xmm\ dataset, while the other instruments suggest a lower value. 
However, in the hard state, we note similar values of the blackbody normalization for different temperatures. The reason why \xmm\ gives a much higher blackbody temperature with respect to \beppo\ and \rxte\ may be due to a contamination from the boundary layer emission visible during the \xmm\ observation and not during the other ones (of course the boundary layer emission may be directly visible when it is not completely comptonized in the corona).

\subsection{The inclination of the system}

The inclination angle of the system with respect to the line of sight, the inner radius of the disk, the emissivity index, and the centroid energy of the line are determined by the profile of the lines \citep{fabian_89}. These parameters are mutually correlated and, it may result difficult to  disentangle their contribution to the overall line shape.
For example, as discussed by \citet{Cackett_10}, the inclination and the emissivity index play a similar role in determining the line profile.
A high value of these parameters makes the line broad and less peaked. So a high value of the inclination and a low value of the emissivity index will give a similar  profile than a low value of the inclination and a high value of the emissivity index.

In the soft state the observed emission lines, and especially the iron line profile, allow us to obtain a good constraint on the inclination angle of the system. Using \textsc{reflionx}, we found $i = 35-40^{\circ}$. 
Applying \textsc{xillver} instead of \textsc{reflionx} to the data, we obtain $i = 25-27^{\circ}$ (the difference between the best fit values of the inclination angle and the ionization parameter obtained with \textsc{xillver} and \textsc{reflionx} are caused by a different energy range used to extrapolate the illuminating flux in the two models, J. Garcia private comm.). 
However using other models on the \xmm\ data, which are the best quality data obtained on \mxb,\ such as \textsc{diskline} and \textsc{pexriv}, the inclination is found at $38-41^{\circ}$, in agreement with \textsc{reflionx}. 
Because the $ \chi^{2}$/dof is higher in the case of \textsc{xillver} and because all the other models indicate the same range of values for the inclination, we conclude that the inclination of \mxb\ is between $35-41^{\circ}$ with respect to the line of sight.

\subsection{Geometry of the accretion disk}

In all the models, the \textsc{rdblur} component was necessary to improve the fit. The mildly relativistic blurring was applied to  the entire reflection spectrum, confirming the common origin of the reflection features in the inner part of the accretion disk, where strong relativistic effects broaden emission and absorption features.
This component gives us information on the inner radius of the accretion disk, $R_\mathrm{in} = 10-16$ $R_\mathrm{g}$ in the soft state, and  $R_\mathrm{in} = 19-59$ $R_\mathrm{g}$ ($R_\mathrm{in} = 26-65$ $R_\mathrm{g}$ for the inclination fixed at $39^{\circ}$) in the hard state.
So we have an indication that the accretion disk is close to the neutron star surface in the soft state and truncated further from the compact object in the hard state.
This is in agreement with \citet{Barret_02} and \citet{Olive_03} who interpreted the transitions from one state to the other one in this source with different truncation radii of the accretion disk. 
This is also a possible interpretation for black hole binaries that show clearer transitions from the soft to the hard state and vice versa \citep[e.g.][]{Done_07}. 


%

The spectral state transitions are also associated  with variations in the overall X-ray luminosity. We calculated the accretion rate in both states using the typical value of the accretion efficiency $\eta = 0.2$, corresponding to a neutron star ($M_\mathrm{NS} = 1.4$ $M_\mathrm{\odot}$ and $R_\mathrm{NS} = 10$ km), and to the bolometric luminosities inferred by our spectral modelling. In the soft state, $\dot M_\mathrm{SS} =
1.6\times 10^{-8}$ $M_\mathrm{\odot}$ $yr^{-1}$, while in the hard state the accretion rate decreases, $\dot M_\mathrm{HS} = 2\times 10^{-9}$ $M_\mathrm{\odot}$ $yr^{-1}$. This difference in the accretion rate is consistent with changes in the flow geometry and hence with a different inner radius of the accretion disk.
The evaporation of the inner part of the accretion disk may lead to a truncated disk at low-mass accretion rates \citep[e.g.][]{Meyer_00} .

So we infer a similar geometry to that proposed for black hole binaries where at low luminosity the accretion disk is truncated \citep{Barrio_03, Done_10}. 
However some differences should be observed between these systems, especially in the soft state, because of the presence of the boundary layer in the case of the neutron star binaries. And in fact we find that in \mxb\ the disk is truncated relatively far from the compact object (at more than 10 $R_g$ both in the soft and in the hard state) as it is inferred by the fact that the observed distortion of the iron line profile is never extreme.

\subsection{The ionization parameter}

Reflection models give an indication on the ionization state of the matter in the inner part of the accretion disk: 
$ \xi = 4 \pi F_\mathrm{X} / n_\mathrm{H}$, where $ F_\mathrm{X} $ is the total illuminating flux (erg cm$^{-2}$ s$^{-1}$) and $n_\mathrm{H}$ is the hydrogen number density.
We note that the matter is much more ionized in the soft state, 
$\xi = 3600$ erg cm s$^{-1}$, 
in comparison with $\xi = 210$ erg cm s$^{-1}$ in the hard state. This is again in agreement with the disk-reflection scenario for a truncated disk and with a lower illuminating flux in the hard state. When the accretion rate is high, the disk penetrates into the hot flow, favoring the interactions between the inner disk and the illuminating flux and resulting in a high ionization and of the matter in the disk and, possibly, in a high reflection amplitude \citep[e.g.][]{Poutanen_97}. On the contrary when the accretion rate is low, the disk is truncated further from the compact object, disk matter is less ionized, and the amount of reflection is intrinsically low \citep[e.g.][]{Barrio_03}. This is consistent with the observed energy of the iron line found in the soft state, $E_\mathrm{Fe} = 6.6-6.7$ keV, and in the hard state, $E_\mathrm{Fe} = 6.4-6.5$ keV \citep[see also][who emphasized the influence of the ionization state of the accretion disk on the iron line profile]{Reis_09}.


\subsection{Overabundance of some elements}

The determination of the abundances of heavy elements is important to infer how they have been originated.
Most heavy elements from Si to Fe originate from explosive nucleosynthesis
in supernovae. Type Ia supernovae, which correspond to the explosion of an
accreting white dwarf in a binary system when its mass becomes superior to
1.4 $M_{\odot}$ via mass transfer, provide mainly Fe whereas core collapse
supernovae (SN II, Ib, IIb) associated to the gravitational collapse of
the iron core of a massive star after successive stages of hydrostatic
burning, provide intermediate elements, from Si to Ca.

In the soft state of \mxb,\ \xmm\ detected emission lines corresponding to S, Ca, Ar and Fe. Using reflection models we have investigated a possible overabundance of some elements with respect to their solar abundance. Applying \textsc{reflionx} we have an indication of an iron overabundance by a factor 2--3, likely responsible for the apparent large edge observed in the soft state \citep{Ross_05}. 
This result is in agreement with \citet{dai_10}, who evaluated the iron overbundance by a factor $\sim$ 3 by using \textsc{refbb}, in addition to an overabundance of S with respect to the other elements (or solar abundance).

In the hard state, the lower statistics does not allow to detect other emission lines than Fe, so it is difficult to estimate any overabundance. Letting this parameter (Fe/sol) free to vary in \textsc{reflionx}, its value is close to 1 and the inner radius of the disk is found at about $R_\mathrm{in} = 19-59$ $R_\mathrm{g}$. When we fix the abundance of iron to a factor 2 with respect to the solar abundance, the inner radius is found at a slightly larger
distance from the neutron star $R_\mathrm{in} = 24-79$ $R_\mathrm{g}$, confirming the evidence of a truncated disk in the hard state. 
It is therefore important to obtain good constraints on the abundance of Fe, and other elements if possible, because this parameter has also a direct effect on the estimate of the inner radius of the disk.

\subsection{Comparison with Cyg X-2 and GX 3+1}

Both Cygnus X-2 and GX 3+1 are bright neutron star LMXBs showing spectral features similar to those observed in \mxb.\

The Z-source Cyg X-2 is one of the rare persistent LMXB whose secondary star is easily observed. It appears to be a high inclination system ($i > 60^\circ$) since short-duration dips were detected in its light curve \citep{Vrtilek_88, Orosz_99}. 
The width of the brightest spectral lines of
Mg XII, Si XIV, S XVI, Fe XXV, and Fe XXVI resolved by Chandra indicate velocity dispersion of the order of 1\,000 to 3\,000 km/s \citep{Schulz_09}, and are consistent with a stationary, dense and hot accretion disk corona. 
Moreover, a \suzaku\ observation revealed the presence of a red-skewed wing of the K$\alpha$ iron line in this source \citep{Shaposhnikov_09}, possibly explained by reflection of X-ray radiation from a cold accretion disk or by Compton down-scattering in a mildly relativistic wind-outflow  \citep{Laurent_07}.
In the case of \mxb,\ the observed emission lines could not be resolved into blending of different lines, nor with \xmm\ \citep{disalvo_09}, nor with the \chandra\ High Energy Transmission Grating \citep{disalvo_05}.  We infer velocities associated with the different emission lines of $\sim 11\,000$ km/s ($E_\mathrm{Fe} = 6.69 \pm 0.01$ keV, $\sigma_\mathrm{Fe} \sim 0.25$ keV; $E_\mathrm{Ar} = 3.31 \pm 0.02$ keV, $\sigma_\mathrm{Ar} \sim 0.13$ keV) and a red-skewed K$\alpha$ iron line, with line width similar to the one observed in Cyg X-2 ($\sigma_\mathrm{Fe} \sim 0.22-0.25$ keV). In both sources, intercombination line dominates the Fe XXV triplet. However, the width measured for the 6.68 keV line is much larger in \mxb\ than in the case of the \chandra\ observation of Cyg X-2, but it is similar to the line width measured in the \suzaku\ observation of Cyg X-2.
Because of the lower statistics of Chandra grating spectra, it may be possible that the red wing of the line is not easily observed by Chandra and is better detected by the large area instruments of \suzaku\ and \xmm.\ 

Another peculiar system presenting very similar features to those observed in \mxb,\ 
is the type-I X-ray burster GX 3+1. Recently, a broad and asymmetric iron line, in addition to Ar XVIII and Ca XIX lines, has been detected with \xmm\ in this persistent and bright atoll source \citep{piraino_12}. 
The iron line profile is well fitted by a relativistically smeared profile and is thought to originate from reflection in the inner parts of the accretion disk. As in \mxb,\ the line profile does not depend upon the photon pile-up fraction in the EPIC-pn spectrum, in agreement with \citet{miller_10}. 
The parameters obtained from the line profile modeled with a gaussian or a diskline, such as the emission line width, the inclination of the system with respect to the line of sight, $35^\circ < i < 44^\circ$, and the inner disk radius, are remarkably similar to those we find for \mxb,\ indicating similar geometry and physical parameters of the disk-corona system in these bright sources.

\section{Conclusions}

Reflection features present complex profiles which depend mainly on the relativistic blurring, caused by Doppler effects due to the high Keplerian velocities and, possibly, gravitational redshift at the inner disk radius,
on the incident flux, on the ionization state of the matter in the disk, on the abundance of the elements and on the inclination of the system with respect to the line of sight. The study of these features gives invaluable information on the system.
We have performed a broad band (0.4--200 keV) and moderately high energy resolution spectral analysis of the X-ray burster \mxb\ both in the soft and in the hard state using data from \xmm,\ \beppo,\ and \rxte\ observatories. This source is particularly interesting since it shows several reflection features observed at a high signal-to-noise ratio. We have fitted these features with several self-consistent reflection models in order to test the common origin of all these features, which are all compatible with being produced by reflection of the primary Comptonization spectrum on the inner accretion disk. In this scenario we have inferred the main parameters of the inner accretion disk. In particular we find the inclination of the system with respect to the line of sight that is constrained in the range $35-41^{\circ}$, the inner radius of the disk which increases from $10-16$ $R_\mathrm{g}$ in the soft state up to $26-65$ $R_\mathrm{g}$ in the hard state, the ionization parameter which decreases from $\sim$ 3600 erg cm s$^{-1}$ in the soft state to 210 erg cm s$^{-1}$ in the hard state. We also find an indication for an iron overabundance with respect to its solar abundance by a factor 2--3. All these results appear to be strong against the particular reflection model used to fit these features and against possible distortion caused by photon pile-up in the XMM-Newton/EPIC-pn CCDs.
 
We have also discussed the differences in the spectral parameters between the soft and the hard state. The results found are consistent with the following scenario.
At low luminosity, the accretion disk is truncated further from the neutron star, 
so the interaction efficiency of the disk photons with the hot electrons of the corona is lower. The rate of photons coming from the disk is also lower because of the cooler temperature of the disk.
This results in a hard spectrum and a low-ionization reflection. At higher luminosity, the mass accretion rate increases and the inner radius of the disk moves closer to the compact object. The soft photons from the disk are much more efficient in cooling the corona, resulting in a softer spectrum. In addition to this, reflection increases due to a stronger irradiation of the disk, and the matter becomes more ionized. 
Moreover, the emission lines are broadened by stronger Doppler effects as the disk approaches the compact object.

This scenario is generally well supported by the timing analysis through power density spectra where correlations are observed between the characteristic frequencies of the fast time variability and the position of the source in the CD or its spectral state \citep[e.g.][]{Olive_03}, with characteristic frequencies increasing with increasing the inferred mass accretion rate.

\begin{acknowledgements}
This work was supported by the Initial Training Network ITN 215212: Black 
Hole Universe funded by the European Community. 
A. Papitto acknowledges the support by the grants AYA2009-07391 and SGR2009-811, as well as the Formosa program TW2010005 and iLINK program 2011-0303.
We thank the referee for valuable comments which helped in improving the manuscript.
\end{acknowledgements}

\bibliographystyle{aa}
\bibliography{biblio_1705}

\begin{table*}[c]
\caption{\xmm\ and \beppo\ observation details}

\label{tabps}
\centering
\renewcommand{\footnoterule}{}  
\begin{tabular}{lcccccc} 
\hline \hline

\\

\textbf{Satellite}& \textbf{Obs. ID} & \textbf{Obs. date} & \textbf{Exp. time (ks)} & \textbf{Count rate ASM (c/s)} & \textbf{State source}\\

   \hline

\xmm\ & 0402300201 & 26/08/2006 & 34.72 & 1 & Hard
\\
            & 0551270201 & 24/08/2008 & 45.17 & 19 & Soft
\\

\beppo\ & 21292001 & 20/08/2000 & 43.5 & 18 & Soft
\\
              &  & 03/10/2000 & 48 & 3 & Hard
\\

\hline
 \\

\end{tabular}
\vskip 0.5cm
\end{table*}

\begin{table*}[c]
\caption{Selected \rxte\ observation details}

\label{tabps:rxtedet}
\centering
\renewcommand{\footnoterule}{}  
\begin{tabular}{ccccc} 
\hline \hline

\\

\textbf{Source state}& \textbf{Obs. ID} & \textbf{Obs. date (MJD)} & \textbf{Count rate (c/s)} & \textbf{Hardness} \\

   \hline

SS & 93060-01-15-00 & 54332.3 & 642.4 & 0.613\\
    & 93060-01-16-00 & 54336.4 & 804.4 & 0.647\\
    & 93060-01-19-01 & 54348.7 & 1049.0 & 0.620\\
    & 93060-01-19-02 & 54349.0 & 1011.0 & 0.621\\
    & 93060-01-25-00 & 54372.5 & 834.5 & 0.630\\
    & 93060-01-76-00 & 54576.5 & 726.9 & 0.620\\
    & 93060-01-84-00 & 54608.5 & 833.7 & 0.628\\
    & 93060-01-91-00 & 54636.8 & 732.4 & 0.625\\
    & 93060-01-95-00 & 54652.6 & 785.0 & 0.617\\
    & 93060-01-01-10 & 54668.1 & 832.2 & 0.646\\
    & 93060-01-02-10 & 54672.3 & 849.8 & 0.637\\
    & 93060-01-07-10 & 54692.5 & 761.5 & 0.614\\
    & 93060-01-10-10 & 54704.4 & 768.4 & 0.632\\
    & 93060-01-12-10 & 54712.2 & 791.4 & 0.617\\
    & 93060-01-19-10 & 54740.3 & 782.5 & 0.631\\
    & 94060-01-20-00 & 54942.5 & 805.5 & 0.609\\
    & 94060-01-22-00 & 54950.6 & 834.7 & 0.622\\
    & 95060-01-71-00 & 55478.6 & 879.0 & 0.620\\

\hline

HS & 91039-01-01-41 & 53541.0 & 79.7 & 0.760 \\
    & 91039-01-01-42 & 53543.0 & 78.1 & 0.749\\
    & 91039-01-01-43 & 53544.8 & 82.6 & 0.751\\
    & 91039-01-01-50 & 53546.0 & 91.3 & 0.767\\
    & 91039-01-02-40 & 53658.3 & 64.9 & 0.736\\
    & 93060-01-07-00 & 54300.4 & 70.6 & 0.747\\
    & 93060-01-52-01 & 54480.6 &  67.1 & 0.746\\
    & 94060-01-08-00 & 54894.9 & 52.1 & 0.738\\
    & 95060-01-19-01 & 55270.5 & 69.6 & 0.749\\
    & 95060-01-33-00 & 55326.4 & 52.5 & 0.733\\
    & 95060-01-46-00 & 55378.7 & 57.7 & 0.750\\

\hline
 
\end{tabular}
\vskip 0.5cm
\end{table*}

\begin{landscape}
\begin{table}[t]
\caption{Evaluation of the pile-up effects on the \xmm\ EPIC-pn data of \mxb\ in the soft state, using the RGS and pn spectra. We compare the values of the parameters we obtain when we exclude 0, 1, 2, until 7 brightest central columns in the pn CCD, which correspond to spectra named pn-all, pn-1, pn-2 until pn-7, respectively, in the table.
The model consists of const*phabs*rdblur*edge*(bbody+compTT+gauss+gauss+gauss+gauss).
The $\sigma$ of the 4 gaussians and the outer radius of the disk are frozen to 0 and 3500 $R_\mathrm{g}$, respectively.
 We also include in the last column the results from \beppo\ data in the soft state as reported by \citet{Piraino_07} to compare the values of the parameters obtained with \xmm\ with those obtained with non-CCD instruments. The model used to fit the \beppo\ spectrum consists of phabs*(bbody+compTT+powerlaw+diskline).}

\label{tabps:pileup7}
\centering
\renewcommand{\footnoterule}{}  
\begin{tabular}{llccccccccc} 
\hline \hline

\\

\textbf{Component}& \textbf{Parameter} & \textbf{pn-all} & \textbf{pn-1} & \textbf{pn-2} & \textbf{pn-3} & \textbf{pn-4} & \textbf{pn-5} & \textbf{pn-6} & \textbf{pn-7} & \textbf{BeppoSAX} \\

   \hline

   phabs & $ N_\mathrm{H} ~ (\times 10^{22}cm^{-2})$ & $1.6 \pm 0.1$ & $1.6 \pm 0.1$ & $1.6 \pm 0.1$ & $1.6 \pm 0.1$ & $1.6 \pm 0.1$ & $1.6 \pm 0.1$ & $1.6 \pm 0.1$ & $1.6 \pm 0.1$ & $1.9 \pm 0.1 $
 \\
   rdblur & Betor  & $-2.30^{+0.04}_{-0.02}$ & -2.32 $\pm 0.03$ & $-2.35^{+0.05}_{-0.04}$ & $-2.39^{+0.06}_{-0.07}$  & $-2.35^{+0.04}_{-0.07} $  & $-2.39^{+0.08}_{-0.07}$ & $-2.4 \pm 0.1$ & $-2.4 \pm 0.1$ & -2 (frozen)
 \\
   rdblur & $ R_\mathrm{in}$ (GM/$ c^{2}$)  & $14 \pm 2$ & $14^{+1}_{-2} $ & $14^{+3}_{-1}$ & $15 \pm 2$ & $15^{+3}_{-2} $  & $16 \pm 3$ & $16^{+4}_{-3}$  & $17^{+4}_{-3}$ & $8^{+4}_{-2} $
 \\
   rdblur & i ($^{\circ}$) & $38 \pm 1$ & 39 $\pm 1$ & $38^{+2}_{-1} $ & $39 \pm 2$  & $41^{+2}_{-1} $ & 42 $\pm 2$ & $40^{+3}_{-2} $ & $40^{+6}_{-3} $ & $28^{+20}_{-8} $
 \\
   edge & E (keV) & $8.6 \pm 0.1$ & $ 8.6 \pm 0.1$ & $ 8.6 \pm 0.1$ & $8.6 \pm 0.1$ & $ 8.7 \pm 0.1$  & $8.7^{+0.2}_{-0.1}$ & $8.8 \pm 0.2$  & $8.6^{+0.3}_{-0.5}$ & -
 \\
   edge & Max $\tau$ $(\times 10^{-2}) $ & $5.1 \pm 0.4$ & $ 5.7 \pm 0.4$ & $ 7 \pm 1$ & $ 7 \pm 1$ & $ 8 \pm 2$ & $ 8 \pm 2$  & $7^{+3}_{-2}$ & $5^{+3}_{-2}$ & -
 \\
   bbody & $ kT_\mathrm{bb} $ (keV) & $0.57 \pm 0.01$ & $0.57 \pm 0.01 $ & $0.57 \pm 0.01 $ & $0.56 \pm 0.01 $ & $0.55 \pm 0.02 $ & $0.55 \pm 0.02 $ & $0.54 \pm 0.01$ & $0.54 \pm 0.02$ & $ 0.56 \pm 0.01 $
 \\
   bbody & Norm $(\times 10^{-2}) $ & $3.0 \pm 0.1 $ & $2.9 \pm 0.1 $ & $3.1 \pm 0.1 $ & $3.0 \pm 0.1 $ & $3.2 \pm 0.1 $ & $3.2 \pm 0.1 $ & $3.4 \pm 0.1$ & $3.4 \pm 0.1$ & $ 2.2 \pm 0.1 $ 
 \\
   compTT & $ kT_\mathrm{o}$ (keV) & $1.31 \pm 0.01$ & $1.29 \pm 0.02$ & $1.30^{+0.02}_{-0.03}$ & $1.34^{+0.02}_{-0.05} $ & $1.33^{+0.03}_{-0.08} $ & $1.35^{+0.05}_{-0.09} $ & $1.36 \pm 0.05$ & $1.39^{+0.06}_{-0.04}$ & $1.13^{+0.05}_{-0.02} $
 \\
   compTT & $ kT_\mathrm{e}$ (keV) & $6^{+19}_{-2}$ & $ 4^{+3}_{-1} $ & $ 4^{+4}_{-1} $  & $4^{+14}_{-1} $& $ 4^{+11}_{-1} $ & $ 4^{+15}_{-2} $ & $4^{+18}_{-1} $ & $4^{+5}_{-1} $ & $ 2.7 \pm 0.1 $
 \\ 
   compTT & $\tau$ & $4.7^{+13}_{-0.1} $ & $ 6.8 \pm 0.1$ & $ 6.6^{+0.1}_{-0.3} $ & $ 5.9 \pm 0.3$ & $ 6^{+7}_{-1} $ & $5.2^{+0.6}_{-0.5}$ & $5.1^{+0.5}_{-0.6}$ & $4^{+60}_{-1}$ & $ 11.0 \pm 0.6 $
 \\
   compTT & Norm $ (\times 10^{-2})$  & $18^{+6}_{-5}$ & $ 29^{+8}_{-2} $ & $ 30 \pm 1 $ & $30^{+65}_{-2} $  & $ 30^{+1}_{-7} $ & $ 30^{+1}_{-12} $  & $33^{+170}_{-12} $ & $32^{+411}_{-8} $ & $ 35 \pm 2 $
 \\
 gauss & E (keV) & $ 2.64 \pm 0.03 $ & $ 2.67 \pm 0.03 $ & $ 2.65 \pm 0.03 $ & $ 2.65 \pm 0.03 $ & $ 2.67 \pm 0.05 $ & $ 2.66 \pm 0.05 $ & $2.7^{+0.9}_{-0.7}$ & $2.8^{+0.1}_{-0.2}$ & -
 \\
  gauss & Norm $ (\times 10^{-3})$ & $0.9 \pm 0.2 $ &  $ 1.3 \pm 0.2 $ &  $ 2.1 \pm 0.3 $ & $1.6^{+0.5}_{-0.3}$ & $ 1.7 \pm 0.6 $  &$1.8^{+0.8}_{-0.7}$  & $1.8^{+0.8}_{-0.9}$ & $1.3^{+0.9}_{-0.7}$ & -
 \\ 
 gauss & E (keV) & $3.29 \pm 0.01$ & $ 3.29^{+0.01}_{-0.02}  $ & $3.29 \pm 0.02$ & $ 3.30^{+0.03}_{-0.02}  $  & $ 3.30^{+0.03}_{-0.02}  $  & $ 3.27^{+0.04}_{-0.02}  $  & $3.27 \pm 0.03$ & $3.28^{+0.03}_{-0.04}$ & -
 \\
  gauss & Norm $ (\times 10^{-3})$ & $1.9 \pm 0.2 $ & $2.1 \pm 0.2  $ & $2.4 \pm 0.2$ & $2.6 \pm 0.3$ & $ 3.2^{+0.3}_{-0.5} $ & $3.1\pm 0.6$  & $ 3.4^{+0.5}_{-0.7} $& $2.8\pm 0.7$  & -
 \\
  gauss & E (keV) & $3.89^{+0.01}_{-0.02}$ & $ 3.87^{+0.04}_{-0.01}  $ & $ 3.89 \pm 0.02 $ & $ 3.88 \pm 0.02 $ & $ 3.88 \pm 0.02 $  & $ 3.86 \pm 0.03 $ & $3.84^{+0.03}_{-0.02}$ & $3.87^{+0.03}_{-0.05}$ & -
 \\
  gauss & Norm $ (\times 10^{-3})$ & $1.7 \pm 0.1$ & $ 2.0 \pm 0.1 $ & $ 2.1 \pm 0.2 $ & $ 2.4 \pm 0.2 $  & $ 2.9 \pm 0.4 $ & $3.0^{+0.5}_{-0.2}$ & $ 3.3 \pm 0.6 $ & $3.1^{+0.9}_{-0.6}$ & -
 \\
  gauss & E (keV) & $6.64 \pm 0.01$ & $ 6.64 \pm 0.01 $ & $ 6.64 \pm 0.01 $ & $ 6.62 \pm 0.02 $ & $ 6.62^{+0.03}_{-0.02} $ & $ 6.59^{+0.02}_{-0.03} $ & $6.58 \pm 0.04$ & $6.58 \pm 0.04$ & $ 6.7^{+0.2}_{-0.5} $
 \\
  gauss & Norm $ (\times 10^{-3})$ & $3.8^{+0.1}_{-0.2}$ & $ 4.1 \pm 0.2 $ & $ 4.3 \pm 0.1 $ & $ 4.1^{+0.3}_{-0.2} $ & $ 4.4 \pm 0.4 $  & $4.7 \pm 0.4$  & $4.9^{+0.5}_{-0.4}$ & $4.6^{+0.6}_{-0.4}$ & $ 4.7^{+2}_{-0.6} $
 \\
&   $ \chi^{2} red $ (d.o.f.) & 1.20 (1024) & 1.15 (1024) & 1.14 (1024) & 1.15 (1024) & 1.13 (1024)  & 1.11 (1024) & 1.13 (1024) & 1.14 (1024) & 1.18 (524)
  \\
   \hline
 
\end{tabular}
\vskip 0.5cm
\end{table}
\end{landscape}

\begin{table*}[t]
\caption{Comparison of three different self-consistent, relativistically smeared reflection models (reflionx, xillver and pexriv) applied on the \xmm,\ \beppo\ and \rxte\ spectra of \mxb\ in the soft state. 
Here we present the best fit parameters we obtain for the \xmm\ spectrum. A few parameters of the continuum got different best fit values for the non-simultaneous \beppo\ and \rxte\ spectra; these differences are detailed in Table~\ref{tabps:SSparam}.
The model is const*phabs*rdblur*edge*(bbody+nthcomp+gauss+gauss+ gauss+gauss+highecut*reflection).
For the pexriv model, we have fixed the disk inclination to $\cos i = 0.78$ and the disk temperature to $10^6$ K.
}

\label{tabps:SStot}
\centering
\renewcommand{\footnoterule}{}  
\begin{tabular}{llccccc} 
\hline \hline

\\

\textbf{Component}& \textbf{Parameter} & \textbf{Reflionx}  & \textbf{Xillver} & \textbf{Pexriv}
\\

   \hline

   phabs & $ N_\mathrm{H} ~ (\times 10^{22}cm^{-2})$ & $ 2.08 \pm 0.02 $ & $ 2.04 \pm 0.01 $ & $ 1.91 \pm 0.02 $ 
 \\
   bbody & $ kT_\mathrm{bb} $ (keV) & $ 0.56 \pm 0.01 $ & $ 0.67^{+0.02}_{-0.01} $ & $ 0.52 \pm 0.01 $
 \\
   bbody & Norm $(\times 10^{-2}) $ & $ 2.58 \pm 0.01 $ &  $ 1.7 \pm 0.01 $ & $ 2.94 \pm 0.01 $
 \\
   nthComp & $ \Gamma $ & $ 2.6 \pm 0.1 $ &  2.6 (frozen) & $ 2.3 \pm 0.1 $
 \\
   nthComp & $ kT_\mathrm{e}$ (keV) & $ 3.0 \pm  0.1 $ &  $ 2.9 \pm  0.1 $ &  $ 2.6 \pm  0.1 $
 \\ 
   nthComp & $ kT_\mathrm{seed}$ (keV) & $ 1.30 \pm  0.02  $ & $ 1.39^{+0.04}_{-0.01} $ & $ 1.18^{+0.03}_{-0.02} $
 \\
   nthComp & Norm  & $ 0.14 \pm 0.01 $ & $ 0.11 \pm 0.01 $  & $ 0.17 \pm 0.01 $
 \\
   rdblur & Betor  & $ -2.1 \pm 0.1 $ & $ -2.1 \pm 0.1 $ & $ -2.3 \pm 0.1 $
 \\
   rdblur & $ R_\mathrm{in}$ (GM/$ c^{2}$)  & $ 13 \pm 3$  & $ 11^{+2}_{-1} $ & $ 15^{+2}_{-3} $
 \\
   rdblur & $ R_\mathrm{out}$ (GM/$ c^{2}$)  &  3500 (frozen) & 3500 (frozen) & 3500 (frozen) 
 \\
   rdblur & i ($^{\circ}$) & $ 37 \pm 2 $ & $ 26 \pm 1 $  & $ 40^{+1}_{-2} $
 \\
   edge & E (keV) & $ 8.7 \pm 0.1$  & $ 8.5 \pm 0.1$ & -
 \\
   edge & Max $\tau$ $(\times 10^{-2}) $ & $ 4.2 \pm 0.6$ & $ 3.4 \pm 0.5$ & -
 \\
 gauss & E (keV) & - & - & 2.6 (frozen)
 \\
  gauss & Norm $ (\times 10^{-3})$ & - & - & $ 1.4 \pm 0.2 $
 \\ 
 gauss & E (keV) & $ 3.31 \pm 0.01 $ & - & $ 3.31^{+0.02}_{-0.03} $
 \\
  gauss & Norm $ (\times 10^{-3})$ & $ 1.4 \pm 0.2  $ & - & $ 1.9 \pm 0.2  $
 \\
  gauss & E (keV) & $ 3.92 \pm 0.02 $ & - & $ 3.88^{+0.01}_{-0.03} $
 \\
  gauss & Norm $ (\times 10^{-3})$ & $ 1.4 \pm 0.2 $ & - & $ 2.0 \pm 0.2 $
 \\
  gauss & E (keV) & - & - & $ 6.63 \pm 0.01 $
 \\
  gauss & Norm $ (\times 10^{-3})$ & - & - & $ 4.1 \pm 0.1 $ 
 \\
  highecut & cutoff$_E$  (keV)  &  0.1 (frozen) & 0.1 (frozen) & -
  \\
  highecut &  fold$_E$ (keV)  & 8.1 (2.7*kT$_\mathrm{e}$ of nthComp) & 7.7 (2.7*kT$_\mathrm{e}$ of nthComp)  & 7.0 (2.7*kT$_\mathrm{e}$ of nthComp) 
  \\
  reflection & $ \Gamma $ & 2.6 (= $\Gamma$ of nthComp) & - & 2.3 (= $\Gamma$ of nthComp)
  \\
  reflection &  Norm  & $ (8 \pm 1) \times 10^{-6} $ & $ (1.5 \pm 0.1) \times 10^{-5}$  & 1.218 (frozen)
  \\
  reflection & Fe/Solar & $ 2.5^{+0.4}_{-0.5}  $& 1 (frozen)  & $1.4^{+0.6}_{-0.1}$
  \\
  reflection & Ar, Ca Abund & 1 (frozen) & $ 1.8^{+0.2}_{-0.3} $ &  1 (frozen)
  \\
  reflection & $ \xi $ (erg cm s$^{-1}$) & $ 3578^{+1184}_{-847} $ & $ 1349^{+31}_{-90}$ & $3081^{+2488}_{-1954} $
  \\
  pexriv & Rel-refl  & - & - & $-1^{+1}_{-0.02}$
 \\
      & Total $ \chi^{2} red $ (d.o.f.) & 1.17 (1573) & 1.26 (1576) & 1.16 (1570)
  \\
   \hline
 
\end{tabular}
\vskip 0.5cm
\end{table*}

\begin{table*}[t]
\caption{Self-consistent, relativistically smeared reflection model (reflionx) applied to the \mxb\ spectra from the three satellites (\xmm,\ \beppo\ and \rxte\ ) when the source was in the hard state; the complete model consists of
const*phabs*(bbody + nthComp + rdblur*highecut*reflionx). 
Here we show the best fit parameters we obtain for the \beppo\ spectrum. A few parameters of the continuum got different best fit values for the non-simultaneous \xmm\ and \rxte\ spectra; these differences are detailed in Table~\ref{tabps:paramHS}. The inclination angle is fixed at $37^{\circ}$, while the Fe/sol ratio is left free to vary or is fixed to 2.
}

\label{tabps:HStot}
\centering
\renewcommand{\footnoterule}{}  
\begin{tabular}{llcccc} 
\hline \hline

\\

\textbf{Component}& \textbf{Parameter} & \textbf{Reflionx (Fe/sol = free)} & \textbf{Reflionx (Fe/sol = 2)}  
\\

   \hline

   phabs & $ N_\mathrm{H} ~ (\times 10^{22}cm^{-2})$ &  $ 1.9 \pm 0.1 $ & $ 1.9 \pm 0.1 $  
 \\
   bbody & $ kT_\mathrm{bb} $ (keV) & $ 0.24^{+0.02}_{-0.03} $ & $ 0.26 \pm 0.02 $
 \\
   bbody & Norm $(\times 10^{-3}) $ & $ 2.0 \pm 0.4 $ & $ 1.7 \pm 0.3 $ 
 \\
   nthComp & $ \Gamma $ & $ 1.84 \pm 0.01 $ & $ 1.83 \pm 0.01 $ 
 \\
   nthComp & $ kT_\mathrm{e}$ (keV) & $ 22^{+2}_{-1} $  & $ 21 \pm 1 $
 \\ 
   nthComp & $ kT_\mathrm{seed}$ (keV) & $ 0.69^{+0.02}_{-0.01} $  & $ 0.70 \pm 0.02 $ 
 \\
   nthComp & Norm $ (\times 10^{-2})$ & $ 3.9^{+0.1}_{-0.2} $ & $ 4.0 \pm 0.2 $  
 \\
   rdblur & Betor  & -3 (frozen)  & -3 (frozen)  
 \\
   rdblur & $ R_\mathrm{in}$ (GM/$ c^{2}$)  & $ 31^{+28}_{-12}$ & $ 39^{+40}_{-15}$   
 \\
   rdblur & $ R_\mathrm{out}$ (GM/$ c^{2}$)  &  3500 (frozen)  &  3500 (frozen) 
 \\
   rdblur & i ($^{\circ}$) & 37 (frozen) & 37 (frozen) 
 \\
  highecut & $ cutoff_E  (keV) $ &  0.1 (frozen) &  0.1 (frozen) 
  \\
  highecut & $ fold_E (keV) $ & 22  (= $kT_\mathrm{e}$ of nthComp) & 21  (= $kT_\mathrm{e}$ of nthComp) 
  \\
   reflionx & Fe/Solar & $ 1.1^{+0.5}_{-0.3}$ & 2 (frozen)
  \\
   reflionx & $ \Gamma $ & 1.84 (= $\Gamma$ of nthComp) & 1.83 (= $\Gamma$ of nthComp) 
  \\
   reflionx & $ \xi $ (erg cm s$^{-1}$) & $  209^{+9}_{-5}$ & $ 204^{+4}_{-3}$
  \\
   reflionx &  Norm $ (\times 10^{-5})$  & $ 1.7 \pm 0.3 $ & $ 1.8^{+0.6}_{-0.3} $  
  \\
   & Total $ \chi^{2} red $ (d.o.f.) & 1.12 (980) & 1.13 (981)
  \\
   \hline
 
\end{tabular}
\vskip 0.5cm
\end{table*}

\begin{figure}[t]
\includegraphics[height=6.5cm]{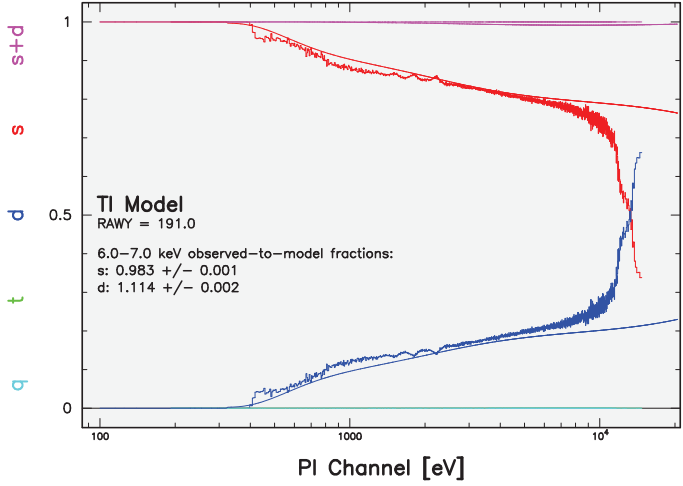}
\caption{Estimation of the pile-up fraction in the EPIC-pn through the \textbf{epatplot} tool at the 6-7 keV iron line energy range when all the columns of the CCD are considered. The plot represents the spectra of the single (red) and double (blue) events. Solid lines indicate the expected fraction from the model curves.}
\label{fig:pileup_all_1}
\end{figure}

\begin{figure}[t]
\includegraphics[height=6.5cm]{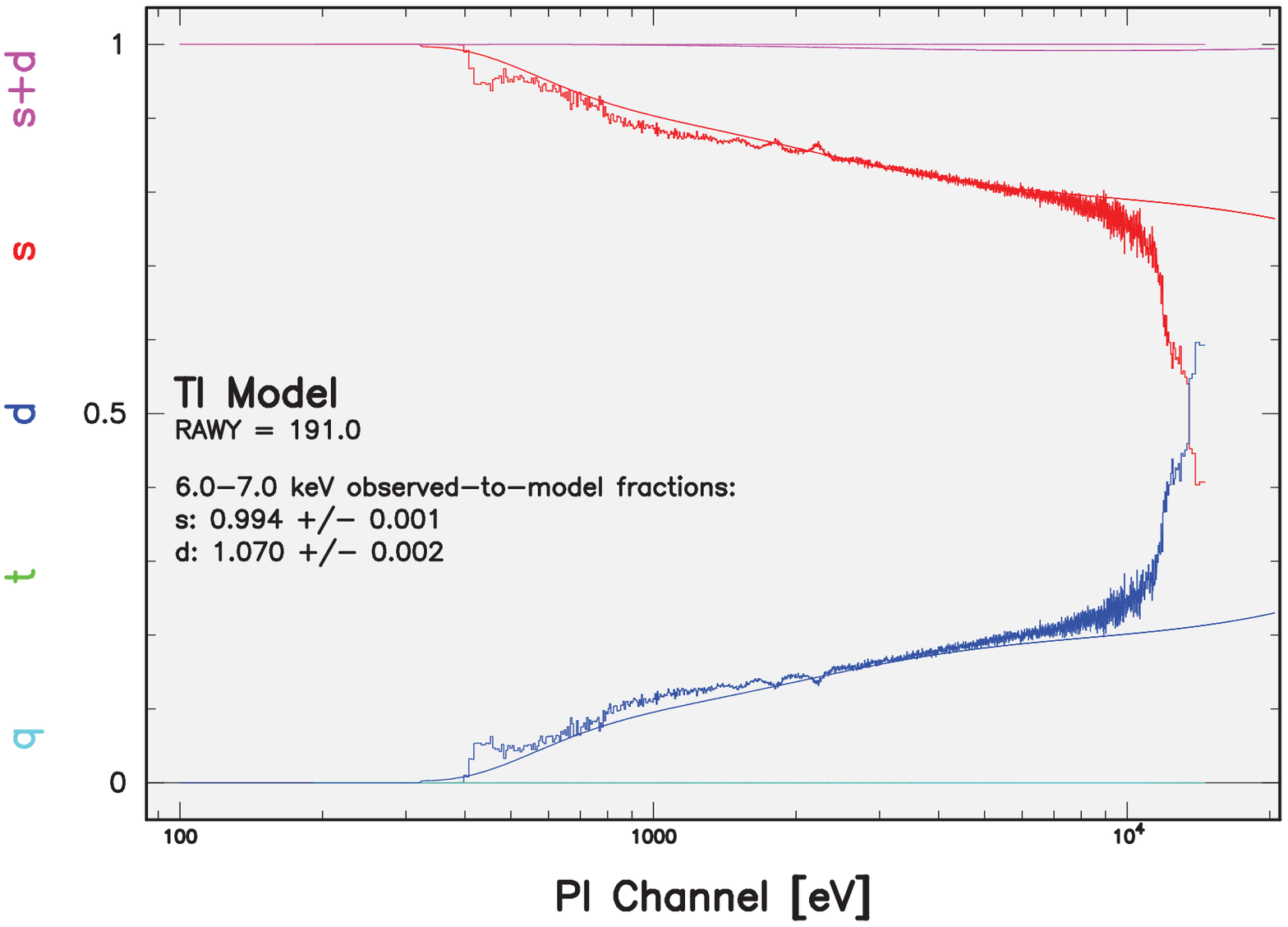}
\caption{Same as in Fig. \ref{fig:pileup_all_1} for pn-1, in which the brightest central column of the CCD has been excluded.}
\label{fig:pileup_all_2}
\end{figure}

\begin{figure}[t]
\includegraphics[height=6.5cm]{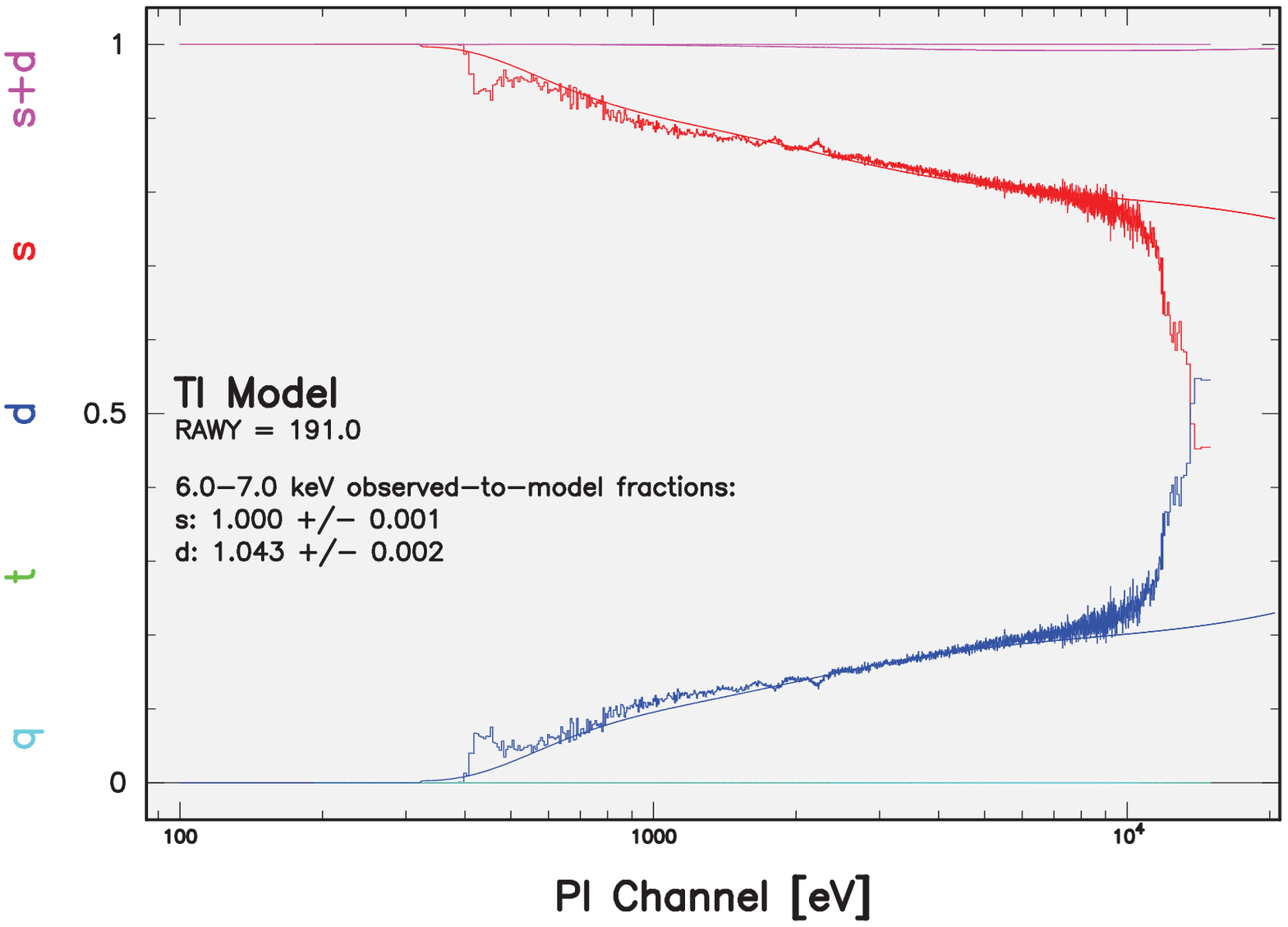}
\caption{Same as in Fig. \ref{fig:pileup_all_1} for pn-2, in which the 2 brightest central columns of the CCD have been excluded.}
\label{fig:pileup_all_3}
\end{figure}

\begin{figure}[t]
\includegraphics[height=6.5cm]{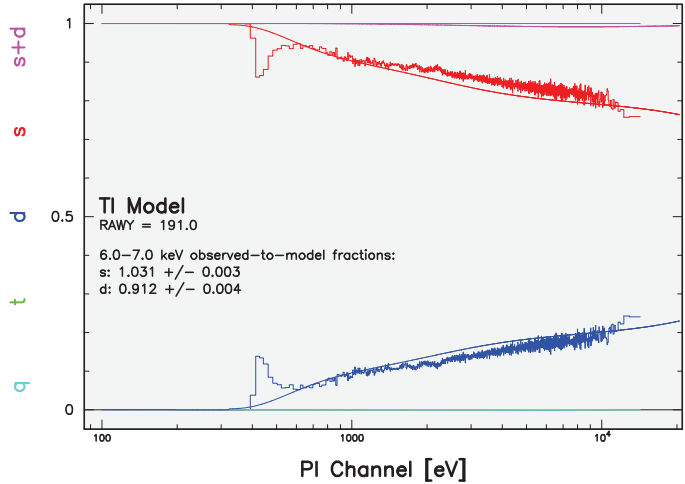}
\caption{Same as in Fig. \ref{fig:pileup_all_1} for pn-7, in which the 7 brightest central columns of the CCD have been excluded.}
\label{fig:pileup_all_4}
\end{figure}

\appendix
\section{Pile-up}

The task \textsc{epatplot} offers the possibility to check whether an observation suffers from pile-up. Here we compare the following different cases, when all the columns of the CCD are used (pn-all), when the brightest central column is excluded (pn-1), when the two brightest central columns are excluded (pn-2), and finally when the 7 brightest columns are excluded (pn-7, as proposed by \citet{Ng_10}) in order to determine if the \mxb\ spectra were affected by significant pile-up during the \xmm\ observation in the soft state.
We note the presence of some pile-up below 2 keV and above 10 keV, unless 7 central columns are excluded. We specify that we have restricted our spectral analysis between 2.4 and 11 keV, which is the range of interest for the study of the iron K$\alpha$ line complex. 
Moreover it is possible to quantify the amount of pile-up in a given energy range. By default this range is from 0.5 to 2 keV which corresponds to the softest part of the spectrum, the most sensitive to pile-up. We estimate the amount of pile-up in the energy band of the iron emission line (6$-$7 keV). 

When we use all the columns of the CCD, we note a deviation between the observed and the expected distribution that testifies the presence of pile-up in the spectrum (Fig.~\ref{fig:pileup_all_1}). The observed to model fractions of the single and double are 0.98 and 1.11 respectively in the 6$-$7 keV band. If we exclude the central column, the deviation is less important, especially for the single events (Fig.~\ref{fig:pileup_all_2}). The observed to model fractions corresponding to the single and double events are 0.99 and 1.07. When we exclude 2 central columns, the single and double distributions follow the expected models meaning that the observation does not suffer anymore of pile-up (Fig.~\ref{fig:pileup_all_3}). The corresponding ratios are 1.00 and 1.04. Finally we exclude the 7 brightest columns. The single and double distributions are overestimated in comparison to the models (Fig.~\ref{fig:pileup_all_4}). The associated ratios are 1.03 and 0.91, respectively, in the 6$-$7 keV energy band.
We therefore conclude that the best solution is to exclude 2 central columns of the CCD in order to avoid pile-up and to have the most correct distribution of single and double events at the iron line energy band. Note also that the range below 2 keV is covered by the RGS, and it is therefore not necessary to eliminate more central columns in the pn CCD with the aim to reduce the pile-up fraction below 2 keV. In other words, we prefer to exclude from our spectral analysis the softer energy range in the pn spectrum, which is covered by the RGS, in order to maximize the statistics of the pn spectrum and the quality of the instrumental response reconstruction in the range of interest for iron line studies.

\end{document}